\documentclass[numberedappendix]{emulateapj}

\usepackage{natbib}
\usepackage{amsmath}

\usepackage{hyperref}
\hypersetup{
    colorlinks=true,
    linkcolor=blue,
    citecolor=blue,
    filecolor=magenta,      
    urlcolor=cyan,
}

\shorttitle{Kinematics of Halo Gas}
\shortauthors{Ho \et}

\begin{document}
%
%
\def\etali{{\it et al.\thinspace}}
\def\etns{{\rm et\thinspace al.}}   

\def\mgIIdb{\ion{Mg}{2} $\lambda\lambda$2796, 2803}
\def\mgIIdbl{\ion{Mg}{2} $\lambda$2796}
\def\mgIIdbu{\ion{Mg}{2} $\lambda$2803}
\def\oI{[\ion{O}{1}] $\lambda$6300}
\def\nII{[\ion{N}{2}] $\lambda\lambda$6548, 6583}
\def\sII{[\ion{S}{2}] $\lambda\lambda$6716, 6731}
\def\oIII{[\ion{O}{3}] $\lambda$5007}

\def\kms{\mbox{km s$^{-1}$}}
\def\kmstb{km s$^{-1}$}

%
%
\def \dlow {\mbox{$400 {\rm ~l~mm}^{-1}$}}
\def \dhigh {\mbox{$600 {\rm ~l~mm}^{-1}$}}
\newcommand{\be}{\begin{equation}} \newcommand{\ba}{\begin{eqnarray}}
\newcommand{\ee}{\end{equation}} \newcommand{\ea}{\end{eqnarray}}
\def\etal{{\it et al.\thinspace}}
\def\-{{\em{---}}}
\def \mA {\mbox{${\rm m \AA} $} }
\def \rr {\mbox{${\rm RR}$} }
\def \rarb {\mbox{${\rm R_AR_B}$} }
\def \rara {\mbox{${\rm R_AR_A}$} }
\def \dd {\mbox{${\rm DD}$} }
\def \dada {\mbox{${\rm D_AD_A}$} }
\def \dadb {\mbox{${\rm D_AD_B}$} }
\def \dr {\mbox{${\rm DR}$} }
\def \darb {\mbox{${\rm D_AR_B}$} }
\def \dara {\mbox{${\rm D_AR_A}$} }
\def \dbra {\mbox{${\rm D_BR_A}$} }
\def \hMpc      {h^{-1}{\rm\ Mpc}}
\def \hkpc      {h^{-1}{\rm\ kpc}}
\def \h         {\hbox{$\, h$} }
\def \hinv      {\hbox{$\, h^{-1}$} }
\def \hinvseven    {\hbox{$\, h_{70}^{-1}$} }
\def\ewr{\mbox {EW$_r$}}
\def\ewo{\mbox {EW$_o$}}
\def\H7{\mbox {$h_{0.7}$}}
\def\naI{\mbox {\ion{Na}{1}}}
\def\mgI{\mbox {\ion{Mg}{1}}}
\def\feI{\mbox {\ion{Fe}{1}}}
\def\oVI{\mbox {\ion{O}{6}}}
\def\znII{\mbox {\sc Zn~II~}}
\def\crII{\mbox {\sc Cr~II~}}
\def\alI{\mbox {\sc Al~I~}}
\def\alII{\mbox {\sc Al~II~}}
\def\alIII{\mbox {\sc Al~III~}}
\def\mgII{\mbox {\ion{Mg}{2}}}
\def\mnII{\mbox {\ion{Mn}{2}}}
\def\niII{\mbox {\ion{Ni}{2}}}
\def\feII{\mbox {\ion{Fe}{2}}}
\def\feIII{\mbox {\ion{Fe}{3}}}
\def\cIV{\mbox {\ion{C}{4}}}
\def\sV{\mbox {\ion{S}{5}}}
\def\siIV{\mbox {\ion{Si}{4}}}
\def\siII{\mbox {\ion{Si}{2}}}
\def\siI{\mbox {\ion{Si}{1}}}
\def\cII{\mbox {\ion{C}{2}}}
\def\cIII{\mbox {\ion{C}{3}}}
\def\llambda{\mbox {$\lambda$}}
\def\mstar{\mbox {$M_*$}}
\def\hlen{\mbox {$h_{0.7}^{-1}$}}
\def\lstarlya{\mbox {$L^*_{Ly\alpha}$}}
\def\IZw18{I~Zw~18}
\def\m82{M82}
\def\Ab{Abell~}
\def\gi{\mbox {\rm g-i}}
\def\ug{\mbox {\rm u-g}}
\def\br{\mbox {\rm b-r}}
\def\eqn{equation}
\def\vesc{\mbox {$v_{\rm esc}$}}
\def\heha{\mbox {He~I~$\lambda 5876$ / H$\alpha$}}
\def\xhe{\mbox {$\chi({\rm He}) / \chi({\rm H})$} }
\def\heii{\mbox {${\rm He}^+$}}
\def\he{\mbox {\rm He}}
\def\hii{\mbox {${\rm H}^+$}}
\def\h{\mbox {\rm H}}
\def\mab{\mbox {$\rm m_{AB}$}}
\def\ssp{\baselineskip=13pt plus 1pt minus 1pt}
\def\tsp{\baselineskip=5pt plus 1pt minus 1pt}
%
%
\def\deg{\mbox {$^{\circ}$}}
\def\msun{\mbox {${\rm ~M_\odot}$}}
\def\zsun{\mbox {${\rm ~Z_{\odot}}$}}
\def\lsun{\mbox {${~\rm L_\odot}$}}
\def\msunyr{\mbox {$~{\rm M_\odot}$~yr$^{-1}$}}
\def\angs{\mbox {~\AA}}
\def\lya{\mbox {Ly$\alpha$}}
\def\lyb{\mbox {Ly$\beta$}}
\def\Ha{\mbox {H$\alpha$}}
\def\Hb{\mbox {H$\beta$}}
\def\Hg{\mbox {H$\gamma$}}
\def\tion{\mbox {$T_{\rm ion}$~}}
\def\ch{\mbox {$\bigtriangleup$}}
\def\grad{\mbox {$\bigtriangledown$}}
\def\lstar{\mbox {$L^*$}}
\def\line{\mbox {~$\lambda$}}
\def\lines{\mbox {~$\lambda\lambda$~}}
\def\h0{\mbox {~H$_0$}}
\def\q0{\mbox {~q$_0$}}
%
%
\def\auroral{[OIII]~$\lambda4363$~}
\def\auroral{[OIII]~$\lambda4363$~}
\def\ohsun{\mbox {(O/H)$_{\odot}$~}}

\def\O1ha{[OI]$\lambda6300$~/~H$\alpha$~}
\def\Ru{[OII]$\lambda\lambda3727$~/~[OIII]$\lambda5007$~}
\def\s2ha{[SII]$\lambda\lambda6717,31$~/~H$\alpha$~}
\def\2z2{HeII~$\lambda4686$~}
\def\z7{[NII]~$\lambda6583$ }
\def\N2{[NII]~$\lambda6583$~/~H$\alpha$~}
\def\16z2{[SII]~$\lambda\lambda6717, 6731$ }
\def\HgI{HgI~$\lambda4358$~}
\def\Sdensity{[SII]~$\lambda6717 / \lambda6731$}
\def\Temp{[OIII]~$\lambda\lambda4959 + 5007 ~{\rm to}~ \lambda4363$~}
%
%
\def\j{J}
\def\n{NGC~}
\def\asec{\ifmmode {'' }\else $''~$\fi}  
\def\amin{\ifmmode {' }\else $'~$\fi}    
\def\arcsper{\ifmmode \rlap.{'' }\else $\rlap{.}'' $\fi} 
\def\arcmper{\ifmmode \rlap.{' }\else $\rlap{.}' $\fi} 
\def\sles{\lesssim}
\def\sgreat{\gtrsim}
%
%
\def\gapp{\mbox {$_>\atop{^\sim}$}}  
\def\lapp{\mbox {$_<\atop{^\sim}$}}  
%
\def\kms{\mbox {~km~s$^{-1}$}}
\def\ergsec{~ergs~s$^{-1}$~}
\def\sb{~ergs~s$^{-1}$~cm$^{-2}$~arcsec$^{-2}$}
\def\flux{~ergs~s$^{-1}$~cm$^{-2}$}
\def\flam{~ergs~s$^{-1}$~cm$^{-2}$ \AA$^{-1}$}
\def\cm3{~cm$^{-3}$}
\def\col{\mbox {~cm$^{-2}$}}
\def\mpc3{~Mpc$^{3}$}
\def\mpc-3{~Mpc$^{-3}$}
\def\rate{~sec$~{-1}$}
\def\um{~${\mu}$m~}
\def\fig{{Figure}}
\def\figs{{Figures}}
\def\tbl{{Table}~}
\def\sec{{Sec.}~}
\def\x{{X-ray}~}
\def\xs{{X-rays}~}
\def\X{{X-Ray}~}

%
\def\et{{\rm et\thinspace al.}\ }   
\def\ets{{\rm et\thinspace al.'s}\ }   
\def\reff{\par\noindent\parskip=1pt\hangindent=3pc\hangafter=1}
%
%

%
\def\beginrefs{
         {\normalsize}
         {\noindent}
         \small
        \baselineskip=11pt
        \parindent=0pt
        \frenchspacing
        \parskip=1pt plus 1pt
        \everypar={\hangindent=0.42in}}

\def\nedit{}
\def\clm{}

\interfootnotelinepenalty=10000

\title{
Quasars Probing Galaxies: I. Signatures of Gas Accretion at Redshift $z \approx 0.2$
	\altaffilmark{$\dagger$} 	\altaffilmark{$\ddagger$}
          }

\author{{Stephanie H. Ho\altaffilmark{1}}, 
{Crystal L. Martin\altaffilmark{1},}
{Glenn G. Kacprzak\altaffilmark{2},}
{Christopher W. Churchill\altaffilmark{3}}}
\email{shho@physics.ucsb.edu,cmartin@physics.ucsb.edu}

\altaffiltext{$\dagger$}{Based on data obtained at the W.M. Keck Observatory, which is operated as a scientific partnership among the California Institute of Technology, the University of California, and the National Aeronautics and Space Administration. The Observatory was made possible by the generous financial support of the W.M. Keck Foundation.}
\altaffiltext{$\ddagger$}{Some of the observations were obtained with the Apache Point Observatory 3.5 meter telescope, 
which is owned and operated by the Astrophysical Research Consortium.}

\altaffiltext{1}{Department of Physics, University of California, Santa Barbara, CA 93106, USA}
\altaffiltext{2}{Centre for Astrophysics and Supercomputing, Swinburne University of Technology, Hawthorn, Victoria 3122, Australia}
\altaffiltext{3}{Department of Astronomy, New Mexico State University, Las Cruces, NM 88003, USA}

\begin{abstract}

We describe the kinematics of circumgalactic gas near the galactic plane, combining
new measurements of galaxy rotation curves and spectroscopy of background quasars.
\clm{The sightlines pass within 19--93 kpc of the target galaxy and generally detect
\mgII\ absorption. 
The \mgII\ Doppler shifts have the same sign as the galactic rotation, so the cold gas
co-rotates with the galaxy.  Because the absorption spans a broader velocity range
than disk rotation can explain, we explore simple models for the circumgalactic kinematics.
Gas spiraling inwards (near the disk plane) offers a successful description of the observations.
}
An Appendix describes the addition of tangential and radial gas flows and
illustrates how the sign of the disk inclination produces testable 
differences in the projected line-of-sight velocity range. 
\clm{
This inflow interpretation implies that cold flow disks remain common down to redshift 
$z \approx 0.2$ and prolong star formation by supplying gas to the disk.
}

\end{abstract}


\section{Introduction}
\label{sec:intro}

Stars and interstellar gas account for only $30$\% of the baryons associated with 
galaxies similar in mass to the Milky Way (MW; \citealt{McGaugh2010}).  The missing baryons 
almost certainly reside in galaxy halos but are challenging to detect in emission due
to their low density. This region between a galaxy and the intergalactic medium is known as the 
circumgalactic medium (CGM, \citealt{Shull2014}).  

Gas accretion from the CGM plays an important role in galaxy evolution.  
A significant rate of infall is required to extend the gas consumption time of
galactic disks and explain the colors of galactic disks along the Hubble sequence
\citep{Kennicutt1998}.
Accretion of gas with lower metallicity than the disk easily 
explains the relative paucity of low metallicity stars in the disk
\citep{vandenBergh1962,SommerLarsen1991,Woolf2012}, a problem not limited to the MW galaxy
\citep{Worthey1996}.  
Remarkably, the cosmological baryonic accretion rate appears to largely 
determine the mass and redshift dependence of the star formation rate 
from $z \approx 2$ to the present \citep{Bouche2010,Dave2012}.

Simple models based on this equilibrium require nearly half the baryons accreted by 
dark matter halos to cycle through galactic disks. The prevalence and speed
of galactic outflows \nedit{\citep{Heckman2000}}, especially among galaxies at intermediate redshift 
\citep{Weiner2009,Erb2012,CLMartin2012,Rubin2014}, 
provide direct evidence for gas recycling. The measured metallicity and 
$\alpha$-element enhancement  of the hot wind (relative to the interstellar medium--ISM)
provides direct evidence for enrichment of outflows by  core-collapse supernovae 
\citep{CLMartin2002}.

Direct observations of the inflowing gas, however, remain sparse \citep{Putman2012}. 
The MW is clearly accreting gas. Complex C \citep{Wakker1999} and 
the Magellanic stream \citep{Fox2014} are examples of accretion, 
but our location inside the Galaxy limits identification of inflow to high velocity
clouds \citep{Zheng2015}. 
Spectral observations identify net inflow 
in roughly 5\% of intermediate redshift galaxies \citep{CLMartin2012,Rubin2012}, a
result consistent with inflow covering a small solid angle. The critical limitation of 
these {\it down-the-barrel} observations is that, much like the MW sightlines, they too
require the Doppler shift of inflowing gas to be distinguished from the velocity dispersion 
of the interstellar medium (ISM); hence,  significant mass flux may be missed.

Observing bright quasars (or galaxies) behind foreground galaxies appears to be one way
to advance our empirical understanding of gas accretion 
\nedit{\citep{Kacprzak2010,Kacprzak2011ApJ,Steidel2002,Bouche2013,
Keeney2013,Nielsen2015,DiamondStanic2016,Bouche2016}}.
The challenge here is in determining whether the intervening absorption produced by the CGM 
can be uniquely associated with the inflowing portion of the flows created by gas recycling.  
A promising strategy for working 
around this ambiguity is to leverage geometrical knowledge about the orientation of a galactic 
disk and the quasar sightline (e.g., \citealt{Gauthier2012,Keeney2013}).
Support for this approach can be found in both recent observations and hydrodynamical simulations.

In hydrodynamical simulations, galactic disks grow by accreting cooling gas from the CGM
\citep{Oppenheimer2010,Brook2012,Shen2012,Ford2014,Muratov2015,Christensen2016}.  As cold streams
fall toward a galaxy, torques generated by the disk align the infalling gas with the pre-exisiting 
disk \citep{Danovich2012,Danovich2015}. The newly accreted gas then forms an extended cold flow 
disk out to large radius, which co-rotates with the galaxy \citep{Stewart2011b,Stewart2013}.
This gas supply prolongs star formation in disks and fuels
starbursts during galaxy interactions.  Feedback from star formation in the form of
supernova explosions, radiation pressure, and cosmic rays drives powerful winds. Much
of this wind ejecta is later re-accreted by the disk, thereby setting up the circulation pattern
\citep{Ford2014,Ford2016}.

In a schematic representation of this circulation, much of the gas accretion 
takes place near the disk plane. Galactic winds, in contrast, break out of the 
ISM perpendicular to the disk plane \nedit{\citep{Veilleux2005}}. 
In principle, this variation in 
the physical origin of the absorbing gas will produce observable signatures
as a function of sightline orientation relative to the galactic disk. 
When a galactic disk is viewed at high inclination, minor axis sightlines will intersect 
extraplanar gas and be sensitive to bipolar outflows. Major axis sightlines, in contrast, 
will  maximize sensitivity to orbits in the disk plane and will not intersect 
biconical outflows. Testing this schematic picture proves difficult.
The published literature includes measurements of thousands of low-ionization-state
absorbers, but the orientation and kinematics of the host galaxy have not typically
been measured \citep{Prochter2006}.

Observations have recently begun, however, to distinguish absorbers based on the azimuthal 
angle of the quasar sightline.\footnote{With the vertex fixed at 
       the galactic nucleus, azimuthal angle, $\alpha$, is measured from the 
       galaxy major axis to the quasar.}
The results largely support the above schematic representation.
First, the large equivalent widths detected along minor axis sightlines, $\alpha = 60\deg$--$90\deg$, 
require a broad range of gas velocity and are thus consistent with a galactic outflow origin
\nedit{\citep{Bordoloi2011,Bordoloi2014,Bouche2012,Kacprzak2012Apj,Keeney2013,Lan2014,Nielsen2015}}. 
Second, the paucity of strong absorbers at intermediate azimuthal angles
\citep{Bouche2012, Kacprzak2012Apj} may reflect the same bimodality measured for
the metallicity distribution of the circumgalactic gas \citep{Lehner2013,Quiret2016}. 
In other words, outflows would create  not only the strong minor axis absorption
but also the most metal-enriched systems. Third, the warm gas, which appears to
be unique to the halos of star-forming galaxies \citep{Tumlinson2011} seems to be
concentrated near the galactic minor axis \citep{Kacprzak2015}.

As part of a larger effort to describe the gas kinematics of the inner CGM of star-forming
galaxies (C.~L.~Martin \etns, in preparation), we present new spectroscopy of both redshift $z \approx 0.2$ 
galaxies and background quasars. In this paper, we discuss the 15 galaxy--quasar pairs most
likely to probe extended gas disks. Our approach has several unique aspects. Whereas
traditional studies detect an intervening absorption system and then search for the host
galaxy {\it ex post facto}, our program probes the CGM of a well-defined galaxy population.
We measure rotation curves for each individual galaxy. The spectroscopy is sensitive
to ionic columns $N(Mg^+) \approx\  2 \times 10^{12}$\col, corresponding to gas columns 
$N(H) \sgreat\ 5 \times 10^{16} (Z / \zsun)^{-1}$\col.

We describe the sample selection, observation, data reduction, and measurements in Section~\ref{sec:data}.  
We present results for the gas kinematics in Section~\ref{sec:results}.  We test simple models 
for the gas kinematics in Section~\ref{sec:cgm} to better understand the dynamics of the halo
gas. We summarize the conclusions in Section~\ref{sec:conclusions}.
Throughout the paper, we adopt the cosmology from \citet{Planck2015XIIIarxiv}, with 
$h = 0.6774$, $\Omega_m = 0.3089$, $\Omega_\Lambda = 0.6911$, $\Omega_b = 0.0486$, 
$\sigma_8 = 0.8159$, and $n_s = 0.9667$.  We use the atomic data in \citet{Morton2003}.  
We quote vacuum wavelengths in the near-UV, but air wavelengths for 
rest-frame optical transitions.

\section{Observations and Data Reduction}
\label{sec:data}

We present the spectroscopy of 15 galaxy--quasar pairs selected from the 
Sloan Digital Sky Survey (SDSS) Data Release 9 (DR9) catalog \citep{Ahn2012}, 
using the Low Resolution Imaging Spectrometer (LRIS; \citealt{Oke1995,Rockosi2010})
with the Cassegrain Atmospheric Dispersion Compensatoron (ADC) 
\citep{Phillips2006} on the Keck I telescope.
The blue sensitivity of LRIS allows us to study the near-UV 
\mgIIdb\ absorption doublet at redshift $z\approx0.2$.   
We acquired additional longslit spectra for a subset of galaxies 
using the Double Imaging Spectrograph (DIS) at the Apache Point Observatory (APO) 
3.5m telescope.\footnote{
  Instrument specifications can be found in the manual written by Robert Lupton, which is available at
  \url{http://www.apo.nmsu.edu/35m_operations/35m_manual/Instruments/DIS/DIS_usage.html\#Lupton_Manual} 
  }
We describe the sample selection criteria in Section \ref{sec:sample}, the
LRIS observations in Section \ref{sec:lris-obs}, and the DIS rotation curves in Section 
\ref{sec:dis-obs}.  For a subset of the sample, we supplemented the SDSS imaging with higher
resolution imaging as described in Section~\ref{sec:imaging-obs}.

\subsection{Sample}
\label{sec:sample}

Our study focuses on galaxies with photometric redshifts in the $0.15 < z < 0.3$ range. 
This redshift selection is a compromise between sensitivity to halo gas and
accurate characterization of galaxy morphology. At lower redshift, the strong, near-UV
resonance lines are not accessible from the largest aperture telescopes. At higher 
redshift, the disk position angle and inclination fit to SDSS 
images \citep{York2000,Eisenstein2011} become increasingly less accurate.

We select highly inclined star-forming galaxies from the SDSS photometry. Our color cut
at $M_u - M_r < 2.0 $ selects primarily late-type galaxies. We required an axis ratio of 
$b/a < 0.71$ in the $r$ band. The resulting disk inclinations are greater than  $i > 43\deg$.\footnote{
     We have applied the Hubble formula \citep{Hubble1926} here with $q_0 = 0.2$.} 
Hence, minor axis sightlines will intersect the disk at very large radii, thereby 
selecting extraplanar gas.
Star-forming galaxies at these redshifts tend to be fainter than the knee in 
the galaxy luminosity function at $M_r^* = -20.44 + 5 \log_{10} h = -21.29 \pm 0.01$ 
\citep{Blanton2003}.

We observed 15 background quasars brighter than $u_{qso} \le\ 18.5$. All these
sightlines pass within 93 kpc of a target galaxy. The smallest impact
parameter is 19 kpc, and the median is 57 kpc. The quasar sightlines are near the disk plane
at  azimuthal angles $\le\ 30^{\mathrm{o}}$. This selection focuses attention on
the subset of \mgII\ absorbers most likely associated with gas accretion onto the galactic
disk. Table~\ref{tb:sdss-info} identifies the galaxy--quasar pairs by their coordinates.
Table~\ref{tb:lris-obs} lists the new observations, which are described further below.

\begin{turnpage}
	\centering
\begin{deluxetable*}{lllllllccc}
\tablecaption{Target information from the SDSS DR9 catalog}
\tabletypesize{\normalsize}
\tablewidth{0pt}
\tablehead{
\colhead{QSO Name} &
\colhead{QSO ID} &
\colhead{QSO R.A.} &
\colhead{QSO Decl.} &
\colhead{Galaxy Name} &
\colhead{Galaxy R.A.} &
\colhead{Galaxy Decl.} &
\colhead{$i$ }&
\colhead{$\theta$ ($''$)} & 
\colhead{$\alpha$ (\deg)} 
\\
\colhead{(1)} &
\colhead{(2)} &
\colhead{(3)} &
\colhead{(4)} &
\colhead{(5)} &
\colhead{(6)} &
\colhead{(7)} &
\colhead{(8)} &
\colhead{(9)} &
\colhead{(10)}
}
\startdata
J084234+565350 & 1237663529724674207 & 08 42 35.00 & +56 53 50.2 & J084235+565358 & 08 42 35.98 & +56 53 58.8 & 60 & 11.8 & 4\\
J084723+254105 & 1237664837539332241 & 08 47 23.56 & +25 41 05.4 & J084725+254104 & 08 47 25.06 & +25 41 04.7 & 52 & 20.2 & 14\\
J085215+171143 & 1237667429553537034 & 08 52 15.34 & +17 11 43.9 & J085215+171137 & 08 52 15.36 & +17 11 37.1 & 63 & 6.8 & 24\\
J091954+291408 & 1237664879949054021 & 09 19 54.28 & +29 14 08.4 & J091954+291345 & 09 19 54.11 & +29 13 45.3 & 66\tablenotemark{a} & 23.1 & 11\tablenotemark{a}\\
J102907+421752 & 1237661851997503514 & 10 29 07.73 & +42 17 52.9 & J102907+421737 & 10 29 07.56 & +42 17 37.6 & 50 & 15.4 & 19\\
J103640+565125 & 1237658302741741626 & 10 36 40.74 & +56 51 26.0 & J103643+565119 & 10 36 43.44 & +56 51 19.0 & 62 & 23.2 & 18\\
J123049+071036 & 1237661971186712663 & 12 30 49.67 & +07 10 37.0 & J123049+071050 & 12 30 49.01 & +07 10 50.6 & 45\tablenotemark{b} & 16.8 & 4\tablenotemark{b}\\
J123317+103538 & 1237658493356671023 & 12 33 17.75 & +10 35 38.2 & J123318+103542 & 12 33 18.80 & +10 35 42.1 & 50 & 16.0 & 3\\
J124601+173156 & 1237668590262616102 & 12 46 01.81 & +17 31 56.5 & J124601+173152 & 12 46 01.75 & +17 31 52.2 & 63 & 4.5 & 11\\
J135522+303324 & 1237665180055765029 & 13 55 22.89 & +30 33 24.8 & J135521+303320 & 13 55 21.20 & +30 33 20.4 & 80\tablenotemark{c} & 22.4 & 2\tablenotemark{c}\\
J135734+254204 & 1237665532246884371 & 13 57 34.41 & +25 42 04.6 & J135733+254205 & 13 57 33.86 & +25 42 05.1 & 45 & 7.5 & 12\\
J142501+382100 & 1237662195073876015 & 14 25 01.46 & +38 21 00.5 & J142459+382113 & 14 24 59.82 & +38 21 13.4 & 60 & 23.2 & 15\\
J154741+343357 & 1237662337327300689 & 15 47 41.88 & +34 33 57.3 & J154741+343350 & 15 47 41.46 & +34 33 50.9 & 58 & 8.2 & 2\\
J160907+441734 & 1237659326563614832 & 16 09 07.45 & +44 17 34.4 & J160906+441721 & 16 09 06.72 & +44 17 21.5 & 72 & 15.2 & 13\\
J160951+353843 & 1237662500544184468 & 16 09 51.81 & +35 38 43.8 & J160951+353838 & 16 09 51.62 & +35 38 38.6 & 72\tablenotemark{d} & 5.7 & 27\tablenotemark{d}
\enddata

\tablenotetext{a}{\small 
	Inclination and azimuthal angles measured from the  $r$-band isophotes in the SDSS image.  
        The catalog measurements have potentially included the tidal arm structure.
	The original SDSS DR9 catalog gives $i = 76^\mathrm{o}$ and $\alpha = 23^\mathrm{o}$.
	}
\tablenotetext{b}{\small 
	Inclination and azimuthal angles measured from the NIRC2 image.  
	The original SDSS DR9 catalog gives $i = 68^\mathrm{o}$ and $\alpha = 19^\mathrm{o}$.
	}
\tablenotetext{c}{\small 
	Inclination and azimuthal angles measured from the NIRC2 image.  
	The original SDSS DR9 catalog gives $i = 58^\mathrm{o}$ and $\alpha = 6^\mathrm{o}$.
	}
\tablenotetext{d}{\small 
	Inclination and azimuthal angles measured from the NIRC2 and GMOS images.  
	The original SDSS DR9 catalog gives $i = 59^\mathrm{o}$ and $\alpha = 15^\mathrm{o}$.
	}
	
\tablecomments{
	\small
	(1) Name of the quasar.  
	(2) SDSS object ID of the quasar.
	(3) Quasar R.A.~(J2000).
	(4) Quasar Decl.~(J2000).
	(5) Name of the galaxy.
	(6) Galaxy R.A.~(J2000).
	(7) Galaxy Decl.~(J2000).
	(8) Inclination of the galactic disk, from SDSS DR9 PhotoObjAll unless otherwise noted.
	\nedit{The statistical error from the $b/a$-ratio in SDSS PhotoObjAll catalog. 
		The typical uncertainty is  $5^\mathrm{o}$  and 
			 does not include the systematic error generated
                         by the point spread function.}
	(9) Angular separation between the galaxy and the quasar.
	\nedit{Angular separation has a typical uncertainty of $\sles\ 0\farcs1$, 
          dominated by the error in determining the center of the galaxy.}
	(10) Azimuthal angle, the angle between the galaxy major axis and the quasar sightline.  
	\nedit{Azimuthal angle has a typical uncertainty of $3^\mathrm{o}$, which is
		dominated by the error in the position angle of the major axis.
		The position angle has been measured in the SDSS $r$-band  image unless
             otherwise noted.}
}
\label{tb:sdss-info}  
\end{deluxetable*}

\end{turnpage}


\subsection{Keck LRIS Observations}
\label{sec:lris-obs}

\begin{center}
\begin{deluxetable*}{llcll}[htb]
\tablecaption{Observations}\tablenotemark{b}
\tabletypesize{\normalsize}
\tablewidth{0pt}
\tablehead{
\colhead{Instrument} &
\colhead{Target} &
\colhead{Exposure Time} &
\colhead{Configuration} &
\colhead{Observing Dates (Semester/Quarter)}
\\
\colhead{} &
\colhead{} &
\colhead{(s)} &
\colhead{}
}

\startdata
Keck/LRIS & J084235+565358/J084234+565350 & 2700/3520 & B1200/R900 & 2015 Mar 21 (2015A)  \\
Gemini/GMOS & J084235+565358/J084234+565350 & 3600 & $r'$ & 2015 Apr 23 (2015A)\tablenotemark{a} \\
Keck/LRIS & J084725+254104/J084723+254105 & 3600/3520 & B1200/R900 & 2015 Mar 22 (2015A) \\
Keck/LRIS & J085215+171137/J085215+171143 & 2700/2640 & B1200/R900 & 2014 May 2 (2014A)\tablenotemark{a}   \\
Gemini/GMOS & J085215+171137/J085215+171143 & 3600 & $r'$ & 2015 Mar 16, 2015 Apr 20 (2015A) \\
Keck/LRIS & J091954+291345/J091954+291408 & 1800/1760 & B1200/R900 & 2015 Mar 22 (2015A)  \\
Keck/LRIS & J102907+421737/J102907+421752 & 2700/2640 & B1200/R900 & 2015 Mar 22 (2015A) \\
APO/DIS   & J102907+421737 & 7200      & B1200/R1200 & 2016 Apr 10 (2016Q2)\tablenotemark{a}\\
Keck/LRIS & J103643+565119/J103640+565125 & 1800/1760 & B1200/R900 & 2015 Mar 22 (2015A)  \\
Keck/LRIS & J123049+071050/J123049+071036 & 3600/3520 & B1200/R900 & 2015 Mar 21 (2015A) \\
Keck/NIRC2 & J123049+071050/J123049+071036 & 600 & $K_s$ & 2015 May 6 (2015A) \\
Keck/LRIS & J123318+103542/J123317+103538 & 1800/1760 & B1200/R900 & 2014 May 3 (2014A)\tablenotemark{a}  \\
Keck/NIRC2 & J123318+103542/J123317+103538 & 600 & $K_s$ & 2015 May 6 (2015A)\\
Keck/LRIS & J124601+173152/J124601+173156 & 2700/2640 & B1200/R900 & 2014 May 3 (2014A)\tablenotemark{a}  \\
Keck/LRIS & J135521+303320/J135522+303324 & 3600/3520 & B1200/R900 & 2014 May 3 (2014A)\tablenotemark{a}  \\
Keck/NIRC2 & J135521+303320/J135522+303324 & 600 & $K_s$ & 2015 May 6 (2015A)\\
Keck/LRIS & J135733+254205/J135734+254204 & 2400/2340 & B1200/R900 & 2014 May 2 (2014A)\tablenotemark{a}  \\
APO/DIS   & J135733+254205 & 3600      & B1200/R1200 & 2015 Mar 25 (2015Q1)\tablenotemark{a}\\
Gemini/GMOS & J135733+254205/J135734+254204 & 3600 & $r'$ & 2015 Apr 23 (2015A)\tablenotemark{a} \\
Keck/LRIS & J142459+382113/J142501+382100 & 1800/1760 & B1200/R900 & 2014 May 2--3 (2015A)\tablenotemark{a}  \\
APO/DIS   & J142459+382113 & 5100      & B1200/R1200 & 2016 Apr 2 (2016Q2)\tablenotemark{a} \\
Keck/NIRC2 & J142459+382113/J142501+382100 & 600 & $K_s$ & 2015 May 6 (2015A)\\
Keck/LRIS & J154741+343350/J154741+343357 & 1800/1760 & B1200/R900 & 2014 May 3 (2014A)\tablenotemark{a}  \\
Keck/NIRC2 & J154741+343350/J154741+343357 & 600 & $K_s$ & 2015 May 6 (2015A)\\
Gemini/GMOS & J154741+343350/J154741+343357 & 3600 & $r'$ & 2015 Mar 23 (2015A)\tablenotemark{a} \\
Keck/LRIS & J160906+441721/J160907+441734 & 5400/5280 & B1200/R900 & 2015 Mar 21--22 (2015A) \\
Gemini/GMOS & J160906+441721/J160907+441734 & 3600 & $r'$ & 2015 Apr 25 (2015A)\\
Keck/LRIS & J160951+353838/J160951+353843 & 1800/1760 & B1200/R900 & 2014 May 2 (2014A)\tablenotemark{a}   \\
Keck/NIRC2 & J160951+353838/J160951+353843 & 600 & $K_s$ & 2015 May 6 (2015A)\\
Gemini/GMOS & J160951+353838/J160951+353843 & 3600 & $r'$ & 2015 Mar 23 (2015A)\tablenotemark{a}
\enddata
\tablenotetext{a}{\small Not photometric.}
\label{tb:lris-obs}  
\end{deluxetable*}

\end{center}
 
We configured the double spectrograph using the D500 dichroic, 
the 1200 lines mm$^{-1}$ grism blazed at 3400 \AA,
and the 900 lines mm$^{-1}$ grating blazed at 5500 \AA. We binned both
detectors $2 \times 2$ producing a spatial scale of $0\farcs27\ \mathrm{pixel}^{-1}$ for 
both LRISb and LRISr, a dispersion of 0.48 \AA\ pixel$^{-1}$ for LRISb, and 
a dispersion of 1.06 \AA\ pixel$^{-1}$ for LRISr. We cut slit masks with
slitlet widths of $1\farcs0$ and lengths adequate to measure the sky spectrum. 
We designed the masks to either include nearby galaxies or to align the slitlet
position angle with the major axis of the target galaxy. If neither of these
multiplexing advantages could be realized, then we observed the galaxy--quasar
pair with a 1\farcs0 wide longslit.

Using IRAF\footnote{IRAF, 
\url{http://iraf.noao.edu},
      is distributed
      by the National Optical Astronomy Observatory, 
      which is operated by the Association of Universities for Research in Astronomy (AURA) 
      under a cooperative agreement with the National Science Foundation.},
we removed fixed pattern noise as follows.
We fit the overscan region of the images and subtracted a bias zeropoint. For blue spectra,
pixel-to-pixel sensitivity variations were calibrated using twilight sky and 
internal deuterium flatfield exposures. Exposures of an internal, halogen lamp were 
used to flatfield the red spectra.

Cosmic rays were easily removed from the blue frames using L.A.Cosmic \citep{vanDokkum2001}.
Because of the higher cosmic ray rate on the red detector, we limited the exposure time of 
individual red frames to 900~s, but the high density of sky lines in the red spectra still made
cosmic ray removal challenging. We created cosmic ray masks for the 2015A red frames
using L.A.Cosmic.  For the 2014A red frames, we made masks by comparing individual frames 
to a median frame. The frames for each object were registered and then
combined using the cosmic ray masks and a sigma-clipping algorithm. 
A two-dimensional error frame was created using the standard deviation of
the image stack at each pixel. 

We fit a dispersion solution to arclamp lines taken immediately before or after
the science frames. We adopted vacuum and air wavelengths, respectively, for
the blue and red frames. The root-mean-square (RMS) residuals were 
0.15 and 0.05 \AA\ for LRISb and LRISr spectra, where the difference reflects
the number of arc lines available.  The object frames were rectified using a 
two-dimensional fit to the dispersion solution. 

We checked our dispersion solutions for each target by
comparing the wavelengths of night sky emission lines to a calibrated sky spectrum from 
the Ultraviolet and Visual Echelle Spectrograph (UVES; \citealt{Hanuschik2003}).  The
largest offsets found were less than 25\kms, and we applied a zeropoint shift to
correct for them. We attribute these shifts to difference in rotator angle 
between the calibration and science exposures.  We then applied a heliocentric 
correction to the spectrum of each target to correct for the earth's seasonal
motion.

We fit the median sky level at each wavelength to the portion of 
the slitlet not covering the galaxy or quasar. A one-dimensional object spectrum and 
a variance spectrum were extracted by summing over columns. We
measured the spectral resolution from arc lines. For the blue spectra,
the resolution was  165\kms\ and 105\kms, respectively, in 2014A and 2015A.  
For the red spectra, the average resolution was 105\kms\ and 75\kms\ in 
2014A and 2015A, respectively. 
Higher resolution spectra would resolve structure in the line profiles. Several
velocity components spread over a 20 to 200\kms\ wide trough are normally required 
to fit metal-line systems \citep{Prochaska1997,Charlton1998,Boksenberg2015}. Our
deconvolution is described in Section~\ref{ssec:compare-vw}.

\subsubsection{Red Spectra: Measuring Emission-line Redshifts}
\label{sec:z-sys}

We determined the galaxy systemic redshift using optical emission lines detected in the LRISr 
spectra.  We measured the redshifted wavelengths of the H$\alpha$ emission line for objects 
observed in 2014A,  together with \oI, \nII, and \sII\ 
for objects observed in 2015A on the integrated one-dimensional galaxy spectra.  
For the highest redshift galaxy, J123049+071050, we measured the redshifted wavelengths of 
H$\beta$ and \oIII\ emission lines.
The redshifts listed in Table~\ref{tab:lris_result} define the systemic velocities.


\begin{turnpage}
	\centering
%
\begin{deluxetable*}{llcccccccc}[htb]
\tablecaption{Measurements from LRIS Spectroscopy}
\tabletypesize{\normalsize}
\tablewidth{0pt}
\tablehead{
\colhead{Quasar} &
\colhead{$z_{gal}$} &
\colhead{$b$} &
\colhead{$W^{2796}_r$} &
\colhead{$W^{\mathrm{Mg\ II}}_r$} &
\colhead{$v_D^{\mathrm{Mg\ II}}$} &
\colhead{$\langle v\rangle^{2796}_\mathrm{W}$} & 
\colhead{$\Delta v_{2796}$} & 
\colhead{$\Delta v_{2796}^{intr}$} &
\colhead{$N(Mg^{+})$}
\\
\colhead{Name} &
\colhead{} &
\colhead{(kpc)} &
\colhead{(\AA)} & 
\colhead{(\AA)} & 
\colhead{(\kmstb)} &
\colhead{(\kmstb)} &
\colhead{(\kmstb)} & 
\colhead{(\kmstb)} & 
\colhead{$(\times10^{12}$ cm$^{-2}$)}
\\
\colhead{(1)} &
\colhead{(2)} &
\colhead{(3)} &
\colhead{(4)} &
\colhead{(5)} & 
\colhead{(6)} & 
\colhead{(7)} &
\colhead{(8)} &
\colhead{(9)} &
\colhead{(10)}
}
\startdata
J084234+565350 & 0.21824\tablenotemark{a} & 43 & $\leq 0.24$ & $\leq 0.35$ & \nodata & \nodata & \nodata & \nodata & \nodata\\
J084723+254105 & 0.19591\tablenotemark{a} & 68 & $0.09^{+0.01}_{-0.01}$ & $0.20^{+0.03}_{-0.02}$ & $-103\pm1$ & $-105^{+9}_{-9}$ & [$-176$, $-44$] & [$-119$, $-101$] & $>5.2$\\
J085215+171143 & 0.16921\tablenotemark{b} & 20 & $1.31^{+0.05}_{-0.04}$ & $2.51^{+0.06}_{-0.07}$ & $-39\pm4$ & $-49^{+5}_{-5}$ & [$-225$, $86$] & [$-135$, $-5$] & $>56$\\
J091954+291408\tablenotemark{c} & 0.23288\tablenotemark{a} & 88 & $0.22^{+0.01}_{-0.01}$ & $0.38^{+0.02}_{-0.03}$ & $-154\pm9$ & $-147^{+5}_{-7}$ & [$-266$, $-8$] & [$-208$, $-66$] & $>7.5$\\
                                & & & $0.52^{+0.02}_{-0.02}$ & $0.87^{+0.03}_{-0.03}$ & $137\pm4$ & $131^{+5}_{-4}$ & [$-51$, $292$] & [$7$, $235$] & $>16$\\
J102907+421752 & 0.26238\tablenotemark{a} & 65 & $0.12^{+0.03}_{-0.03}$ & $0.25^{+0.03}_{-0.03}$ & $-53\pm15$ & $-53^{+17}_{-20}$ & [$-134$, $34$] & [$-76$, $-24$] & $>5.8$\\
J103640+565125 & 0.13629\tablenotemark{a,d} & 58 & $0.28^{+0.06}_{-0.08}$ & $0.37^{+0.07}_{-0.10}$ & $412\pm22$ & $392^{+16}_{-22}$ & [$284$, $516$] & [$341$, $459$] & 6.6\tablenotemark{e}\\

J123049+071036 & 0.39946\tablenotemark{f} & 93 & $0.08^{+0.02}_{-0.01}$ & $0.12^{+0.02}_{-0.02}$ & $-91\pm34$\tablenotemark{g} & $-93^{+36}_{-38}$\tablenotemark{g} & [$-172$, $-21$] & [$-115$, $-79$] & $>2.2$\\
J123317+103538 & 0.21040\tablenotemark{b} & 57 & $0.33^{+0.06}_{-0.02}$ & $0.64^{+0.06}_{-0.05}$ & $22\pm9$ & $20^{+8}_{-29}$ & [$-86$, $129$] & [$5$, $39$] & $>14$\\
J124601+173156\tablenotemark{c} & 0.26897\tablenotemark{b} & 19 & $0.31^{+0.03}_{-0.03}$ & $0.58^{+0.04}_{-0.04}$ & $-299\pm11$ & $-293^{+9}_{-8}$ & [$-404$, $-200$] & [$-314$, $-290$] & $>13$\\
J135522+303324 & 0.20690\tablenotemark{b} & 78 & $\leq 0.18$ & $\leq 0.19$ & \nodata & \nodata & \nodata & \nodata & \nodata\\
J135734+254204 & 0.25995\tablenotemark{b} & 31 & $1.49^{+0.02}_{-0.02}$ & $2.65^{+0.02}_{-0.02}$ & $-6\pm1$ & $-33^{+3}_{-3}$ & [$-423$, $197$] & [$-332$, $107$] & $>55$\\
J142501+382100 & 0.21295\tablenotemark{b} & 83 & $0.24^{+0.03}_{-0.02}$ & $0.36^{+0.03}_{-0.03}$ & $9\pm7$ & $21^{+12}_{-14}$ & [$-128$, $221$] & [$-37$, $130$] & 5.7\tablenotemark{e}\\
J154741+343357 & 0.18392\tablenotemark{b} & 26 & $1.01^{+0.02}_{-0.03}$ & $1.91^{+0.04}_{-0.03}$ & $-15\pm2$ & $-12^{+3}_{-3}$ & [$-152$, $156$] & [$-62$, $65$] & $>42$\\
J160907+441734 & 0.14732\tablenotemark{a} & 40 & $0.29^{+0.02}_{-0.02}$ & $0.59^{+0.04}_{-0.03}$ & $-60\pm4$ & $-59^{+5}_{-6}$ & [$-151$, $30$] & [$-94$, $27$] & $>14$\\
J160951+353843 & 0.28940\tablenotemark{b} & 26 & $2.17^{+0.04}_{-0.03}$ & $4.24^{+0.06}_{-0.06}$ & $6\pm2$ & $6^{+2}_{-3}$ & [$-213$, $231$] & [$-122$, $141$] & $>97$
\enddata
\tablenotetext{a}{\small H$\alpha$, \oI, \nII, and \sII\
	emission lines are used to determine the systemic redshift.}
\tablenotetext{b}{\small Only the H$\alpha$ emission line is used to determine the systemic redshift.}
\tablenotetext{c}{\small 
	\nedit{Potential misidentification of host galaxy of the \ion{Mg}{2} absorption.}
	}
\tablenotetext{d}{\small \clm{The blue spectrum shows hints of a second, weaker component just below the detection limit.}}
\tablenotetext{e}{\small 
	The doublet ratio is consistent with 2:1 within uncertainties, i.e., an optically thin absorption.  
	\nedit{$N(Mg^{+})$ is calculated from the equivalent width of \mgIIdbl\ (i.e., $W^{2796}_r$).}
	}
\tablenotetext{f}{\small H$\beta$ and \oIII\ emission lines are used to determine the systemic redshift.}
\tablenotetext{g}{\small
        The absorption system falls in a part of the spectrum without arclamp lines. Extrapolation of the dispersion solution
        adds an additional error term of 34\kms\ on top of the measurement error.
        }
\tablecomments{
	\small
	(1) Name of the quasar.
	(2) Galaxy systemic redshift measured from emission lines. 	\nedit{Number of significant figures listed for galaxy systemic redshift reflects its uncertainty.}
	(3) Sightline impact parameter.
	(4) Rest-frame equivalent width of \mgIIdbl; $2\sigma$ upper limit is listed for non-detections.  
	(5) Rest-frame equivalent width of both \mgIIdb; $2\sigma$ upper limit is listed for non-detections.
	(6) \ion{Mg}{2} Doppler shift measured from line profile fitting.
	(7) Equivalent width weighted mean \ion{Mg}{2} velocity shift.
	(8) Measured velocity range. 
	(9) Intrinsic velocity range.
	(10) Column density of $\mathrm{Mg}^{+}$. 	\nedit{Lower limit on column density $N(Mg^{+})$, which we derive for optically thin absorption (and unity
          covering fraction).  The doublet
          ratio indicates that many systems are actually optically thick, so the actual columns may be orders of magnitude
          larger in some cases.
	} 
}
\label{tab:lris_result}  
\end{deluxetable*}

\end{turnpage}

\subsubsection{Blue Spectra: Measuring the \mgII\ Absorption}
\label{sec:eqw}

To quantitatively describe the \mgII\ strength and kinematics without introducing a particular 
parametric description, we perturbed each 1D spectrum 1000 times adding
a random value in the range of $[-\sigma,+\sigma]$ at each pixel, 
where $\sigma^2$ is the variance in the flux density.
For each fake spectrum, we identified lines by searching for several
consecutive pixels with flux density at least $1\sigma$ below the continuum level in each
pixel. 
\clm{
This velocity range defines a bandpass for each line. Integration
over these pixels determined the total equivalent width and the 
mean velocity, which we weighted by equivalent width. 
}
When this criterion did not identify a line, we defined a bandpass at the systemic velocity with a width
of three times a resolution element and calculated the equivalent width weighted
mean velocity and an upper limit on the equivalent width.
We took the median of the probability distributions for
the equivalent width, equivalent width weighted mean velocity,  
and the total velocity range as our best estimate of each parameter.
The error bars represent the 68\% confidence interval.

We considered both transitions in the doublet to objectively determine whether \ion{Mg}{2}\  was detected. 
Our doublet detections have a total equivalent width with a significance of at least $3\sigma$; the
wavelength separation of the two lines and their equivalent width ratio must also be consistent with the
\mgII\ transitions.
For the detected systems, 
Table~\ref{tab:lris_result} lists the rest-frame equivalent width $W_r$, 
mean velocity $\langle v\rangle_\mathrm{W}$ 
(weighted by equivalent width),
and velocity width $\Delta v$.

\clm{
As a point of comparison, we also fit the \ion{Mg}{2}\ absorption doublet. We used
the custom software described in \citet{CLMartin2012}, which applies the 
Levenberg--Marquardt algorithm \citep{Press1992} to minimize the chi-squared fit statistic.
Convolution of the intrinsic line profile with a Gaussian model of the line-spread function 
leaves the fitted profiles nearly Gaussian in form. The errors on the fitted Doppler shift are 
obtained from the covariance matrix.  As shown in Table~\ref{tab:lris_result}, the fitted Doppler 
shifts and the equivalent widths of the line profiles generally agree with the direct integration
method described above. We therefore conclude that the results are not sensitive to how we measured 
the Doppler shift.
}

\clm{
We constrain the Mg$^+$ column density assuming optically thin absorption,
which provides a robust lower limit. Our curve-of-growth analysis provides 
higher column densities. We do not list these doublet-ratio results 
in Table~\ref{tab:lris_result} because the errors are poorly constrained. The 
LRIS spectral resolution does not distinguish individual absorption components.
Unresolved, saturated absorption components would produce misleading results.
Along some sightlines, the actual ionic column densities may be orders of magnitude  
higher than our lower limits.
}

\subsection{APO DIS Observations}
\label{sec:dis-obs}

Longslit spectra were obtained using the APO/DIS in 2015 March and 2016 April 
for three galaxies, for which the details are listed in Table~\ref{tb:lris-obs}.  
We used the grating with 1200 lines mm$^{-1}$ for both the blue and red channels, 
which are blazed at 4400 \AA\ and 7300 \AA\ respectively.  
We observed with the default 1 by 1 binning, 
resulting in a spatial scale of 
$0\farcs42\ \mathrm{pixel}^{-1}$ and $0\farcs40\ \mathrm{pixel}^{-1}$
and a wavelength dispersion of 0.62 \AA\ pixel$^{-1}$ and 0.58 \AA\ pixel$^{-1}$
for the blue and red spectra respectively.
We observed each of the three galaxies with a $1\farcs5$ longslit, 
and we aligned the slitlet with the galaxy major axis. 
In this paper, we describe the red spectra, which captured the H$\alpha$ emission.

\clm{The APO data reduction followed standard steps.
The red frames were corrected for the zeropoint bias level, trimmed,
and flat fielded using exposures of an  internal quartz lamp.
Cosmic rays were removed when combining frames. We masked cosmic
rays using L.A. Cosmic \citep{vanDokkum2001}, sometimes editing
the masks around strong sky lines. To clean up a few remainig cosmic
rays, we applied a sigma-clipping algorithm  when combining frames. 
A two-dimensional error frame was also created along with each combined science spectrum, 
which was the standard deviation of each pixel among different exposures.
}

We used a list of vaccum wavelengths for the arclamp calibration.
The RMS error in the dispersion solutions is 0.04 \AA. 
Comparison to a UVES  sky spectrum \citep{Hanuschik2003}, 
converted into vacuum wavelength using the Edl{\'e}n formula \citep{Edlen1966} and 
smoothed to the DIS spectral resolution, showed shifts in the emission lines
up to 0.9 \AA. We applied this correction and then corrected to a heliocentric
reference frame. The spectral resolution of the red spectra was determined from the arclamps' lines, 
and was measured to be 50\kms.

\subsubsection{Rotation Curves}
\label{ssec:PV-method}

To obtain the position--velocity (PV) maps, 
we fit a Gaussian function to the emission-line profile at different spatial positions
on the two-dimensional LRISr and DIS spectra.\footnote{   
       We generally used the extended H$\alpha$ emission line to construct the PV maps. 
       For J123049+071050, we used H$\beta$ since H$\alpha$ is not covered; 
       and since the H$\alpha$ emission line is badly contaminated by a strong skyline  
       in the J160951+353838 spectrum, we measured [\ion{N}{2}] $\lambda$6583.}
We did not typically obtain a DIS spectrum when the LRIS slit was within 30\deg\
of the major axis.
We calculated the projected rotation velocity relative to the redshift derived
from the integrated spectrum.  
We plot the PV diagram of each galaxy in the middle column in Figure \ref{fig:sdss_pv}.

We produce galaxy rotation curves by 
deprojecting the position and velocity from the direct measurements onto 
circular orbits in the disk plane.  
Let $\rho$ be the observed projected distance along the slit with respect to the 
galaxy center, and $v_{los}$ be the line-of-sight velocity detected along the slit.
Following \citet{Chen2005}, if the slitlet is at an angle $\zeta$ from the galaxy major axis, 
we translate the projected distance $\rho$ into galactocentric radius $R$ by 
\begin{equation}
	R = \rho \sqrt{1 + \sin^{2}\zeta \tan^{2}i}
	\label{eq:r-deproj}
\end{equation}
and find the rotational velocity $v_{rot}(R)$ along the disk by 
\begin{equation}
	v_{rot}(R) = \frac{\sqrt{1 + \sin^{2}\zeta \tan^{2}i}}{\sin i \cos \zeta} v_{los}(\rho),
	\label{eq:v-deproj}
\end{equation}
where  $i$ is the galaxy inclination.  
\nedit{
To calculate the error in rotation velocity $v_{rot}(R)$,
we first estimate the uncertainty of velocity measured along the slit.
We perturb each 2D spectrum by adding 
a random value within [$-\sigma,\sigma$] at each pixel, 
where $\sigma^2$ represents the variance of flux density.  
With 1000 fake spectra, 
we obtain the line centroid and its uncertainty 
from the median and 68\% confidence interval 
of the probability distribution.  
We then combine this uncertainty 
with the errors in inclination angle
and position angle of the galaxy major axis from Table~\ref{tb:sdss-info} 
to obtain the errors in deprojected velocities.
}

\subsection{NIRC2 and GMOS Imaging Observations}
\label{sec:imaging-obs}

In this paper, we use new, higher resolution imaging to check for 
measurement errors in the disk position angle and inclination. 
Table~\ref{tb:sdss-info} denotes corrections to the values
tabulated in the SDSS DR9 photoObj catalog.
We introduce the key properties of the imaging
observations here and defer details to a future paper describing the
morphological structure of these galaxies. The imaging observations
are summarized in Table~\ref{tb:lris-obs}.

We obtained images of half the target galaxies  with the NIRC2 camera on the Keck II telescope 
on 2015 May 6.  The quasar was used as the tip-tilt reference for the 
Laser Guide Star Adaptive Optics \citep{vanDam2006,Wizinowich2006} system.
The $K_s$ broadband filter with a central wavelength at 2.146 $\mu$m and a 
bandbass width of 0.311 $\mu$m.  
We used the wide field camera with a field of view of $40'' \times 40''$ 
and a plate scale of $0\farcs04\ \mathrm{pixel}^{-1}$. The median width of the quasar
profile was 0\farcs13 FWHM. We reduced these images
using the data reduction pipeline provided by the  
UCLA/Galactic center group \citep{Ghez2008}.

Images of a few galaxies were obtained in queue mode with the
the Gemini Multi-Object Spectrograph (GMOS, Program GN-2015A-Q-60).
The GMOS science images and calibration data, including the bias and twilight flats, 
were retreived from the 
Gemini Observatory Archive\footnote{\url{https://archive.gemini.edu/}}.  
All data were reduced using the Gemini IRAF Package.  
We used the $r'$-band filter with an effective wavelength of 6300 \AA, and 
wavelength interval between 5620 and 6980 \AA.
The pixel scale was $0\farcs0728$ pixel$^{-1}$, and the spatial resolution
ranged from 0\farcs5 to 0\farcs8 FWHM.

\subsubsection{Galaxy Morphology}

If the morphology of the target galaxy was clearly different from 
that in the SDSS catalog, we fit a galaxy surface brightness profile model.  
We wrote custom software to fit the point-spread function (PSF) of the quasar using a Moffat profile.  
\clm{
To create a model, we convolved an exponential surface brightness profile (SB) with this PSF.  
We perturbed the galaxy image by adding a random value within [$-\sigma$,$\sigma$] 
at each pixel, where $\sigma^2$ is the variance in flux density. 
}
We \nedit{then} find the underlying galaxy profile 
by minimizing the $\chi^2$ between the convolved model and the \nedit{fake} galaxy image.  
\nedit{Iterating this process by 1000 times, 
we obtain the `true' galaxy profile and the uncertainty of each parameter
using the median and 68\% confidence interval.
}
The uncertainties in the fitted inclination and the galaxy major-axis position angle
are usually within $5^\mathrm{o}$ and $3^\mathrm{o}$ respectively.

\subsubsection{\nedit{Galaxy Interactions and Environment}}

\clm{Galaxy group environment affects properties of \ion{Mg}{2} absorption.  
Around group members, \ion{Mg}{2} absorption is detected to larger impact parameters 
than around isolated galaxies \citep{Bordoloi2011}.  
Our target galaxies are generally isolated. We searched each field
for galaxies with photometric redshifts within three standard deviations
of our primary target. In only \nedit{four} of the 15
fields do we find a galaxy as bright as the target within the virial radius.
We describe these \nedit{four} environments here; see the images in Figure~\ref{fig:sdss_pv}.
}

\clm{
\nedit{\underline{\j084723+254105 field.}}
The photometric redshift of the red galaxy labeled {\it FIL}
agrees with the redshift of our primary target.  
The LRIS spectrum of {\it FIL}
shows no emission lines and does not cover the best
spectral region for fitting an absorption-line redshift.
The only other potential group member is at twice the impact
parameter of our primary pair. These three galaxies may
form a group.
}

\clm{
\nedit{\underline{\j091954+291408 field.}}
The bright, red galaxy 4\farcs4 \nedit{northeast} of our target galaxy has a photometric
redshift consistent with our target.  A fainter, red galaxy is located
8\farcs7 south of our target with a similar photometric redshift.   These
three galaxies likely form a group.
}

\clm{
\nedit{\underline{\j102907+421752 field.}}
The galaxy labeled {\it FIL} has spectroscopic redshift, $z = 0.18496$, that
precludes group membership with our target. This field includes another bright 
galaxy, however, which is not shown in the postage stamp because it lies three
times farther from the quasar than our target galaxy. Our target and this more
distant galaxy could be members of the same group.
}

\clm{
\nedit{\underline{\j124601+173156 field.}}
Within the virial radius of our primary galaxy, we find
several galaxies with photometric redshifts consistent with group memberships.  
Slightly further away, we identified another set of galaxies with photometric redshifts also 
consistent with group membership.  
It seems likely that our target galaxy belongs to one of these groups.  
}

\clm{
We also looked for signs of galaxy interactions.
One galaxy, \j160951+353838, shows
a tail-like structure toward the northeast side.  The irregularity
of rotation curve in this region confirms the identification of a merger.
The quasar sightline shows the strongest \ion{Mg}{2} absorption in 
the sample, and we suggest that this is related directly to the merger.
SDSS images of \j124601+173152 show two surface brightness peaks, which
higher resolution imaging may resolve into a merger.
}

\subsubsection{Stellar Mass}
\label{sec:SMHM}

The stellar mass is obtained from spectral energy distribution (SED) fitting with FAST \citep{Kriek2009}, 
using the stellar population synthesis model from \citet{Bruzual2003}, 
Chabrier IMF \citep{Chabrier2003}, and the Calzetti dust extinction law \citep{Calzetti2000}.  
SDSS {\it ugriz} photometry is used for every galaxy. 
\clm{
Supplementary photometry from \textit{Galaxy Evolution Explorer} 
(\textit{GALEX}; \citealt{CMartin2005}) imaging
is listed in Table~\ref{tb:mass} when available.
}
\textit{GALEX} data products \citep{Morrissey2007} are downloaded from MAST\footnote{\url{http://archive.stsci.edu}}, 
and the deepest available image for each field is used for flux measurement.  
\clm{
The same aperture is used in both near-UV and far-UV images.}
A $3\sigma$ upper limit is used when there is no obvious UV emission detected from the target galaxy, 
or when it is badly contaminated by emission from the quasar or other nearby objects.  
All fluxes, i.e., both SDSS and \textit{GALEX}, are then corrected for 
Galactic dust extinction before input into FAST.  
This is done by using the reddening law from \citet{Fitzpatrick1999}, 
together with the $E(B-V)$ retrieved from the 
NASA/IPAC Infrared Science Archive\footnote{\url{http://irsa.ipac.caltech.edu}} 
using the calibration from \citet{Schlafly2011}.  
The stellar mass estimates range from 
$\log (M_*/M_\odot) = 9.4$ to 10.6 with a median of $M_*  = 6.61 \times 10^9 M_\odot$, and 
Table~\ref{tb:mass} lists the individual values for reference.

\begin{center}
\begin{deluxetable*}{lrccccccc}[htb]
\tablecaption{Galaxy Properties}
\tabletypesize{\normalsize}
\tablewidth{0pt}
\tablehead{
\colhead{Galaxy} &
\colhead{NUV/FUV} &
\colhead{$m_{\textrm{NUV}}$\tablenotemark{a}} &
\colhead{$m_{\textrm{FUV}}$\tablenotemark{a}} &
\colhead{$\log({M_\star/{M_\odot})}$\tablenotemark{b}} &
\colhead{$\log({M_{vir}/{M_\odot})}$\tablenotemark{c}} &
\colhead{$r_{vir}$\tablenotemark{d}} &
\colhead{$R/r_{vir}$\tablenotemark{e}} & 
\colhead{$v_{rot}$\tablenotemark{f}}
\\
\colhead{Name} &
\colhead{Survey\tablenotemark{a}} &
\colhead{(AB  mag)} &
\colhead{(AB  mag)} &
\colhead{} &
\colhead{} &
\colhead{(kpc)} &
\colhead{} &
\colhead{(\kmstb)}
}
\startdata
J084235+565358 & AIS/AIS & $\geq22.62$     & $\geq23.30$        & $9.74^{+0.21}_{-0.18}$ & $11.47^{+0.12}_{-0.09}$ & $155^{+15}_{-10}$ & 0.279 & 110\\
J084725+254104 & MIS/MIS & $\geq24.23$     & $\geq25.27$        & $9.80^{+0.20}_{-0.13}$ & $11.50^{+0.12}_{-0.07}$ & $160^{+15}_{-8}$ & 0.444 & 115\\
J085215+171137 & AIS/AIS & $\geq22.11$     & $\geq23.66$        & $9.71^{+0.07}_{-0.13}$ & $11.45^{+0.03}_{-0.07}$ & $156^{+4}_{-7}$ & 0.166\tablenotemark{g} & 150\tablenotemark{g}\\
J091954+291345 & DIS/AIS & $21.20\pm0.03$  & $\geq22.94$     & $10.54^{+0.18}_{-0.17}$ & $12.18^{+0.30}_{-0.22}$ & $264^{+69}_{-41}$ & 0.366 & 260\\
J102907+421737 & AIS/AIS & $\geq22.50$     & $\geq23.15$        & $10.03^{+0.11}_{-0.09}$ & $11.66^{+0.08}_{-0.06}$ & $174^{+12}_{-8}$ & 0.399 & 155\tablenotemark{h}\\
J103643+565119 & DIS/DIS & $20.97\pm0.02$  & $21.40\pm0.05$  & $9.94^{+0.19}_{-0.06}$ & $11.56^{+0.13}_{-0.03}$ & $174^{+18}_{-5}$ & 0.385 & 205\\
J123049+071050 & GII/GII & $22.79\pm0.16$  & $\geq25.01$     & $10.41^{+0.03}_{-0.05}$\tablenotemark{i} & $12.07^{+0.03}_{-0.06}$\tablenotemark{i} & $220^{+6}_{-10}$\tablenotemark{i} & 0.423 & 180\\
J123318+103542 & GII/AIS & $21.96\pm0.11$  & $\geq22.77$     & $9.62^{+0.13}_{-0.11}$ & $11.41^{+0.07}_{-0.05}$ & $148^{+8}_{-6}$ & 0.383 & 170\\
J124601+173152 & GII/AIS & $\geq23.11$     & $\geq22.60$        & $9.54^{+0.02}_{-0.16}$ & $11.38^{+0.01}_{-0.07}$ & $140^{+1}_{-8}$ & 0.146 & 60\\
J135521+303320 & AIS/AIS & $\geq21.43$     & $\geq22.66$        & $9.82^{+0.13}_{-0.11}$ & $11.51^{+0.08}_{-0.06}$ & $161^{+10}_{-7}$ & 0.494 & 155\\
J135733+254205 & 
AIS/AIS & $\geq22.15$     & $\geq23.51$        & $10.24^{+0.09}_{-0.08}$ & $11.83^{+0.09}_{-0.07}$ & $199^{+14}_{-11}$ & 0.159 & 160\tablenotemark{j}\\
J142459+382113 & MIS/AIS & $21.17\pm0.10$  & $\geq22.84$     & $10.27^{+0.18}_{-0.14}$ & $11.84^{+0.20}_{-0.12}$ & $206^{+35}_{-19}$ & 0.441 & 185\tablenotemark{h}\\
J154741+343350 & AIS/AIS & $\geq22.18$     & $\geq24.38$        & $9.81^{+0.17}_{-0.09}$ & $11.50^{+0.10}_{-0.05}$ & $161^{+13}_{-6}$ & 0.161 & 175\\
J160906+441721 & NIL/NIL & \nodata         & \nodata                & $9.44^{+0.20}_{-0.04}$ & $11.32^{+0.09}_{-0.02}$ & $143^{+10}_{-2}$ & 0.347 & 130\\
J160951+353838 & AIS/AIS & $19.34\pm0.08$  & $20.88\pm0.29$  & $10.02^{+0.12}_{-0.12}$ & $11.66^{+0.09}_{-0.08}$ & $172^{+13}_{-10}$ & 0.255 & 50\tablenotemark{k}
\enddata
\tablenotetext{a}{\small AIS: All-sky Imaging Survey.  MIS: Medium Imaging Survey.
        DIS: Deep Imaging Survey.  GII: Guest Investigator Program.  NIL: No available images.
        \clm{
        Fluxes are listed as measured, prior to correction for Galactic reddening.}
        In the absence of a detection, we list the $3\sigma$ limit. }
\tablenotetext{b}{\small Stellar mass and the 68\% confidence levels are obtained 
	from SED fitting using FAST \citep{Kriek2009}.}
\tablenotetext{c}{\small Halo mass is obtained from the stellar mass using the 
	stellar mass--halo mass (SM--HM) relation from \citet{Behroozi2010}.  
	The confidence levels of halo mass are calculated from those of stellar mass.  
	Uncertainties due to the scatters in SM--HM relation are not included.}
\tablenotetext{d}{\small Virial radius is calculated using Eqn.~\ref{eq:Mvir} and the 
	redshift of each galaxy in Table~\ref{tab:lris_result}, 
	expressed in kiloparsecs.  
	The confidence levels of virial radius are calculated from those of the halo mass.  
	}
\tablenotetext{e}{\small $R$ is the radial distance between the sightline and the galaxy center 
	{\it on the disk plane}.
	}
\tablenotetext{f}{
	\small The asymptotic rotation speed $v_{rot}$ has a systematic error of $\sim 20$\kms\ 
	due to the uncertainty in the galaxy systemic velocity.  
	Less extended rotation curves also impose larger uncertainties onto $v_{rot}$.
	}
\tablenotetext{g}{
	\nedit{The GMOS images of the galaxy may suggest a 5\deg\ and sub-degree change in 
		the inclination and major-axis position angle, respectively, from the SDSS measurements.  
		These potential adjustments lead to extra uncertainties on $R/r_{vir}$ and $v_{rot}$
		of no more than 15\%. 
		Furthermore, due of the scatter of individual galaxies around the mean SM--HM relation,
		the uncertainty in $R/r_{vir}$ is dominated by that of $r_{vir}$.
		}
	}
\tablenotetext{h}{
	\small The asymptotic rotation speed $v_{rot}$ is derived using the rotation curve from the APO/DIS spectra.  
	The offset of the Keck/LRIS slitlet from the galaxy major axis exceeds 30\deg.  
	}
\tablenotetext{i}{\small The `galaxy' in the SDSS is resolved into two objects.  
	The SDSS photometry is therefore overestimated which leads to a poor SED fit using FAST.  
	The fitted stellar mass and thereby the halo mass are subjected to systematic errors.}
\tablenotetext{j}{
	\small The listed $v_{rot}$ is the average $v_{rot}$ from the rotation curves derived from the Keck/LRIS and APO/DIS spectra.  
	We have no preference between the two rotation curves since the Keck/LRIS slitlet 
	is only $12$\deg\ from the galaxy major axis.
	}
\tablenotetext{k}{
	\small The rotation curve deviates from disk-like rotation, which leads to a large uncertainty in the 
	asymptotic rotation speed $v_{rot}$.  
	The tail-like structure toward the northeast side of the galaxy 
	suggests the possibility of galaxy interaction.
	}
\label{tb:mass}  
\end{deluxetable*}

\end{center}

\section{General Kinematic Properties of the CGM}
\label{sec:results}
\label{sec:general}

Figure~\ref{fig:sdss_pv} shows the fields of the target galaxies,
the measurements of galactic rotation, and the foreground
\mgII\ absorption in the quasar spectrum.  To facilitate a comparison
of the galactic and circumgalactic gas kinematics,
we have rotated the images to align the LRIS slitlet with the spatial
axis of the PV diagram. We chose the orientation
that places the quasar sightline to the right of the target galaxy.
Since we will argue that modeling the velocity width presents a 
key challenge to understanding the physical origin of the absorption, we 
illustrate the instrumental broadening in all the figures. In Figure~\ref{fig:sdss_pv},
the instrumental resolution of LRIS broadens the intrinsic \mgII\ line profiles
as illustrated by the cyan bars in Figure~\ref{fig:sdss_pv}; the orange bars
represent the instrinsic line widths, 
for which we discuss the resolution correction in Section~\ref{ssec:compare-vw}. 
Several results follow directly from the measurements shown in this figure.


\begin{figure*}
	\centering
	\includegraphics[width=0.98\linewidth]{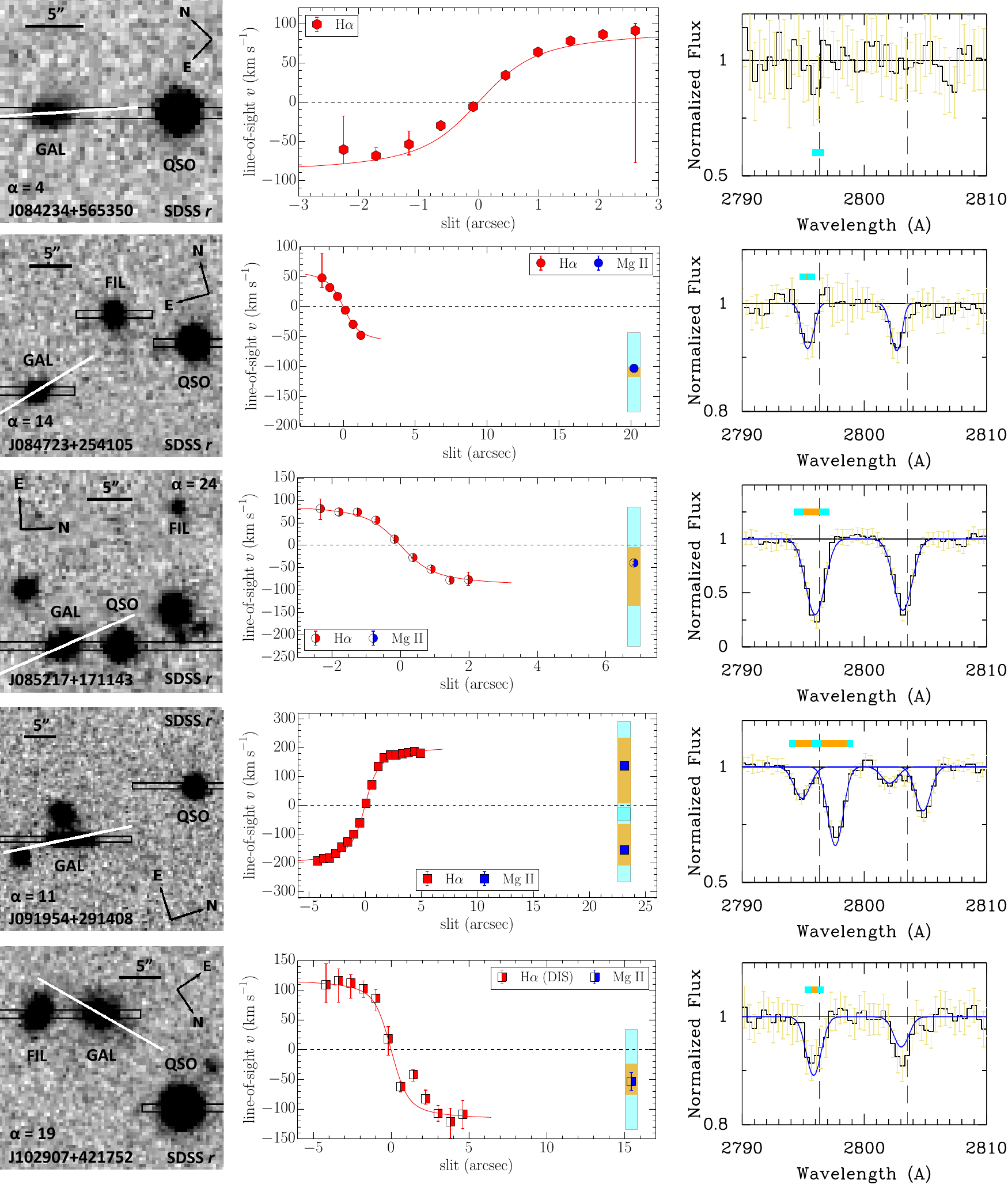}
	\caption{
	Imaging and spectroscopy of galaxy--quasar pairs.
	{\it Left Column:} SDSS $r$-band, GMOS $r$-band image, or NIRC2 $K_s$ band image.
	Labels indicate the quasar, target galaxy, and filler galaxies (if applicable). 
	The black rectangle illustrates the $1\farcs0$ wide LRIS slitlet, and the white
	line indicates the position angle of the galaxy major axis.  
	The $1\farcs5$ wide APO/DIS slitlet was aligned with the galaxy major axis.
	{\it Middle Column:}
	position--velocity diagram for the optical emission-line flux from LRISr and/or DIS.
	We show the line-of-sight velocity by direct measurements (red symbols).  
	{\it Right Column:}
	LRISb spectrum of the \mgII\ system relative to the systemic velocity of the target galaxy. 
	We superimpose the fitted line profile onto the spectrum (blue line) 
	and show the Doppler shift of the \ion{Mg}{2}\ absorption in the quasar sightline 
	on the position-velocity diagram (blue symbol).
	The cyan bar shows the full line-of-sight velocity range detected while the orange
	bar shows the instrinsic width corrected for resolution.
	In cases when no \ion{Mg}{2} absorption is detected, 
	the cyan bar shows the FWHM of a resolution element.  
	For the J135734+254204 sightline, we show the position-velocity map 
	of the galaxy obtained from both the LRIS and DIS spectra.  The LRIS slitlet is 12\deg\
	from the galaxy major axis and we have no preference between the two sets of measurements.
	}
	\label{fig:sdss_pv}
\end{figure*}

\begin{figure*}
	\centering
	\figurenum{1}
	\includegraphics[width=0.98\linewidth]{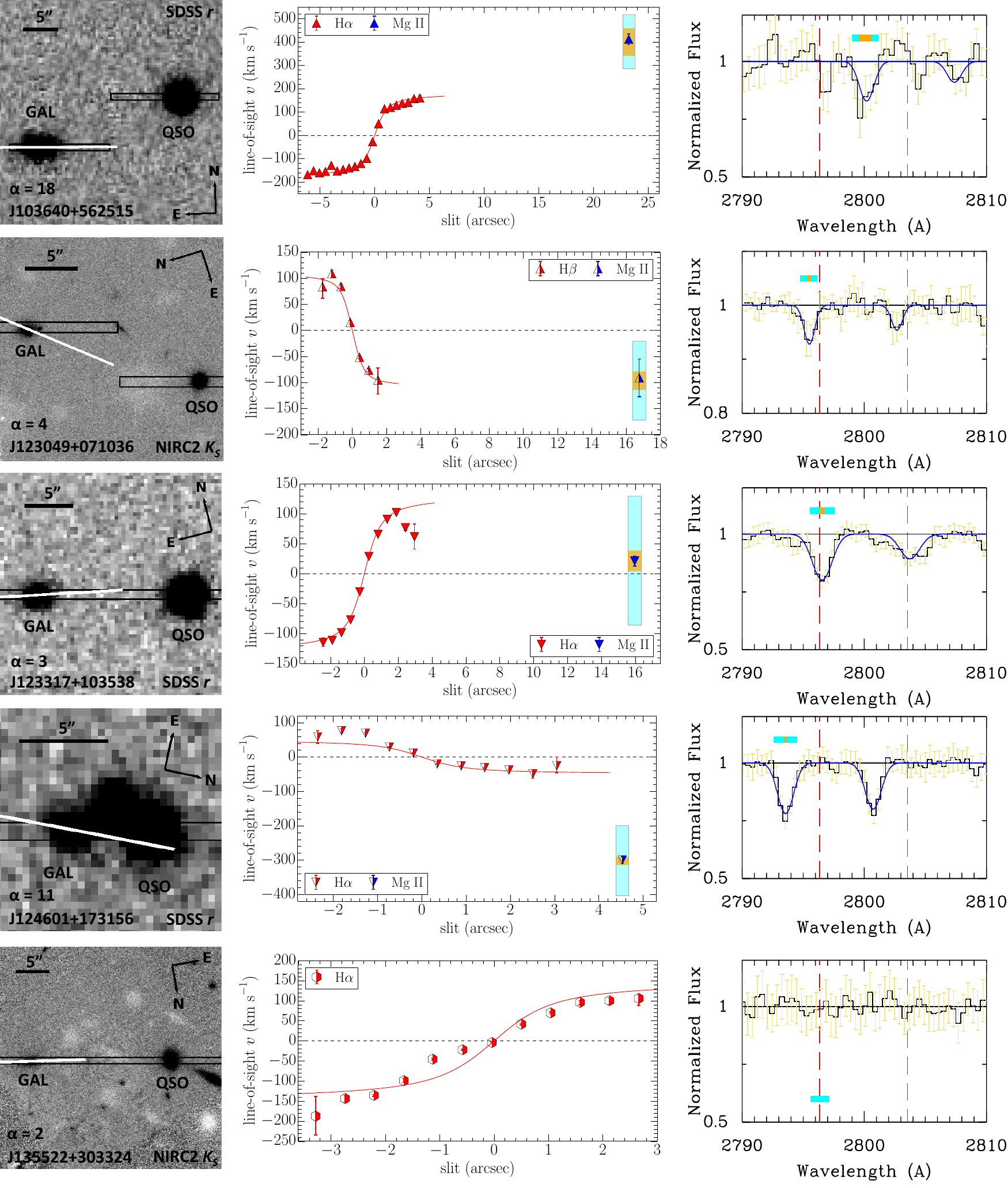}
	\caption{{\it (Continued)}
			}
\end{figure*}

\begin{figure*}
	\centering
	\figurenum{1}
	\includegraphics[width=0.98\linewidth]{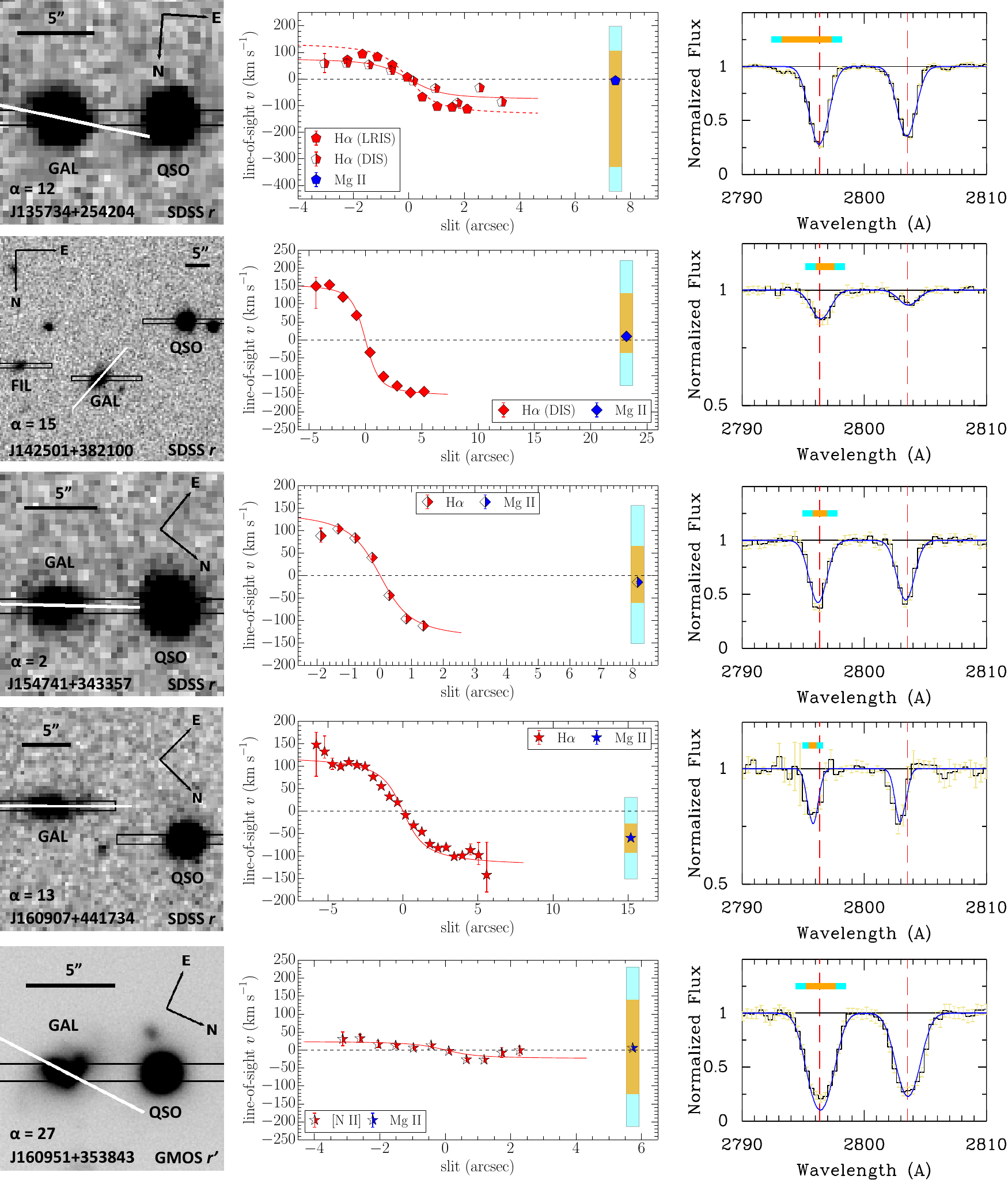}
	\caption{{\it (Continued)}
			}
\end{figure*}

\begin{figure}[htb]
	\centering
	\includegraphics[width=0.85\linewidth]{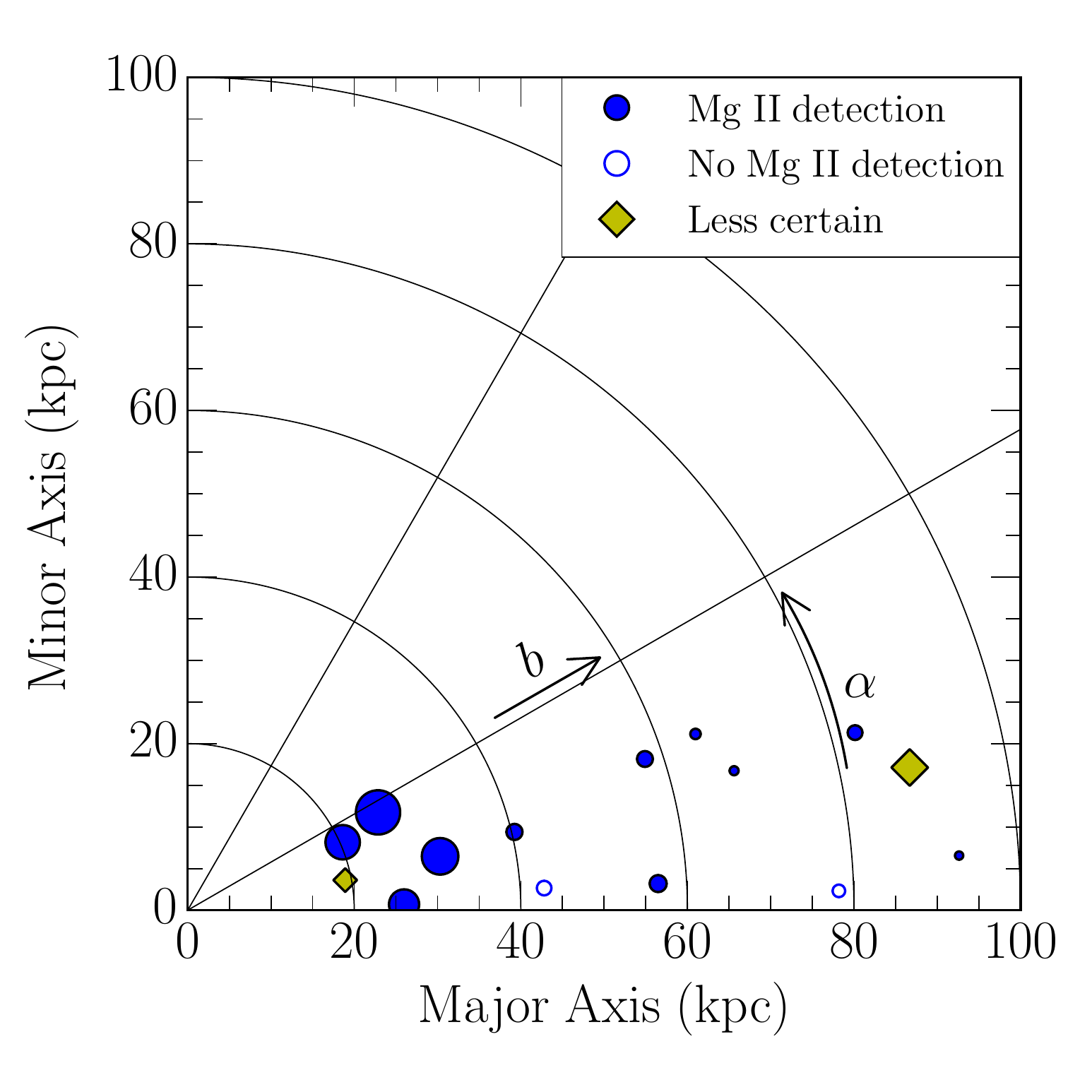}
	\caption{Ensemble of sightlines.  The LRISb spectra detect \mgII\
         absorption at impact parameters from 19 to 93 kpc near the galaxy major axis.  
         \nedit{All measured equivalent widths and upper limits lie within 
		 one standard deviation from the mean equivalent width -- impact parameter		 
		 relations.
		 }
		 The apparent decrease in the covering fraction with impact parameter 
         is \nedit{therefore} largely a sensitivity issue related to the decrease in mean absorption
         strength with the impact parameter. 
		 \nedit{
		 The filled and unfilled markers represent detections and non-detections respectively.  
		 The yellow diamonds flag the uncertain systems as described in the text.  
		 Each marker size scales with the rest-frame equivalent width $W_r^{2796}$ (detection)
		 or its $2\sigma$ upper limit (non-detection).}
	}
	\label{fig:b-alpha} 
\end{figure}

\clm{
We detected \mgII\ absorption in 13 of the 15 quasar sightlines. Most of these
sightlines probe the CGM of isolated galaxies. 
Figure \ref{fig:b-alpha} illustrates the location and strength of the absorption
relative to  the major axis.  The strongest absorber is the merger, \j160951+353838.
}

\clm{
Throughout the manuscript, the yellow diamonds identify two systems with 
uncertain host galaxy assignments. 
The \j091954+291408 sightline intersects a group, 
see Section~2.4.2, and our spectrum detects two absorption systems. 
The statistical relationships between absorber and galaxy properties 
\citep{Lan2014} suggest that our target has a 50\% chance of being the source
of either system. The bright, red galaxy and our target have essentially
the same impact parameter ($b = 88 $~kpc), and red and blue galaxies
are equally likely to produce the measured absorption strengths
($W_r(2796) = 0.52$ and 0.22 \AA).
We assign the stronger system to the blue galaxy (our target) and flag
the weaker system as potentially mis-assigned, i.e., not produced by  our target.
\nedit{
We also flag the \j124601+173156 sightline through the richer group 
that includes galaxy \j124601+173152.
}
\cite{Bordoloi2011} found that overlap of the circumgalactic media of 
group galaxies flattens the radial decline in absorption strength. This 
sightline intersects our target galaxy at $0.14 r_{vir}$, and the
contribution from the CGM of the other likely members (at $r > 0.70 r_{vir}$)
is likely small. Our primary concern is
the object between the quasar and the target galaxy.  
\nedit{
SDSS DR9 classifies this object as a star.
}
We have yet to spectroscopically \nedit{verify} its stellar nature, and an identification
as a compact galaxy would change our interpretation.
}

\subsection{Evidence for Rotation}
\label{sec:rotation}

We \nedit{explore the} connection between the CGM kinematics and the
angular momentum vector of the galactic disk.  
\nedit{
We detect 14 \mgII\ systems in 13 sightlines.  
We exclude the two flagged systems.  
Since only one of the two systems along the \j091954+291408 sightline is excluded, 
12 systems in 12 sightlines remain in our analysis.
Among these 12 sightlines,
}
the Doppler shift of the \mgII\ absorption is 
offset more than 20\kms\ from the galaxy velocity in eight sightlines. 
Inspection of Figure~\ref{fig:sdss_pv} shows that the sign of the \mgII\ 
Doppler shift always matches the sign of the galactic rotation on the quasar
side of the major axis. 

Contrast this result with expections for clouds on random orbits in 
a spherical halo. The shot noise from individual clouds would align 
the Doppler shifts half the time and produce anti-alignments in all other sightlines.
Taking the probability of alignment in any one sightline as $p = 0.50$,  
the chance of finding eight alignments among 12 sightlines is just 12\%.

Consideration of the systems without a net Doppler shift further
lowers the odds of obtaining our data from a random, spherical cloud distribution.
Our measurement errors on the systemic velocities for four systems -- \j135734+254204, 
\j142501+382100, \j15471+343357, and \j160951+352843 -- are comparable
to the uncertainties in the net Doppler shift of the \mgII\ absorption. 
It seems unlikely that higher S/N ratio data would reveal anti-correlations
in all four of these systems. 
The true number of (projected) angular momentum alignments among our 12 sightlines 
very likely exceeds eight.

Our work was motivated in part by the pioneering study of \citet{Steidel2002} who
published rotation curves for the hosts of five \mgII\ absorbers at $z \approx 0.5$.
Three of their systems had azimuthal angles and disk inclinations consistent with 
our selection critiera, and the Doppler shifts of all three of these \mgII\ systems share 
the sign of the galactic rotation.  
With alignments now detected in 11 of our combined 15 sightlines, 
we can conclude that the circumgalactic gas clouds do not follow random
orbits. The data require a component of angular momentum that appears to be aligned
with the gas disk of the galaxy. A qualitatively similar trend, though quantitatively less
significant,  has also been seen at larger azimuthal angles \citep{Kacprzak2010,Kacprzak2011ApJ};
and we will quantify the dependence on azimuthal angle in future work.

\subsection{Comparison to Galaxy Masses}

In our study, the absorption systems span a very broad velocity range compared to 
the thermal line width. 
We interpret the broad line widths as evidence that many clouds contribute to each 
absorption system. The resulting picture requires a population of clouds to produce
each \mgII\ system. If the circumgalactic clouds follow random orbits in a roughly
spherical halo, for example, then the line widths would reflect the depth of the
gravitational potential while the Doppler shifts would be near the systemic velocity.
The shot noise resulting from the finite number of clouds could produce net
Doppler shifts along single sightlines, but no average shift would be measured.

To compare the velocity range of the \mgII\ absorption to the motion of clouds
in virial equilibrium, we estimated the halo mass and virial radius 
of each galaxy. We adopted a stellar mass--halo mass relation derived from 
abundance matching  by \citeauthor{Behroozi2010}~(\citeyear{Behroozi2010}, hereafter B10).
We obtained a  median halo mass of $\log(M_{vir}/M_\odot) = 11.51$, which is significantly
lower than the canonical quenching mass of $10^{12}$\msun
\citep{Conroy2009,Yang2012,Kravtsov2014}.
We note that the scatter of individual galaxies around the mean stellar mass--halo mass
relation introduces an uncertainty of $\pm 0.30$~dex on any single halo mass estimate.
This error is in addition to the statistical errors listed in Table~\ref{tb:mass}.

We define the virial radius for each halo by
\begin{equation}
r_{vir} = M_{vir}(z)^{1/3}  \left( \frac{4\pi}{3} \Delta_{vir}(z)\rho_c(z) \right)^{-1/3}
	~.
	\label{eq:Mvir}
\end{equation}
The B10 definition of the virial mass follows \citet{Bryan1998} who define 
the overdensity $\Delta_{vir}(z)$ with respect to the critical density at redshift $z$ 
by the expression
\begin{align}
	\Delta_{vir} &= 18 \pi^2 + 82x - 39 x^2
	~,\\
	x = \Omega(z) - 1 & \quad \textrm{and} \quad \Omega(z) = \rho_m(z)/\rho_c(z)
	~,
	\label{eq:delta-vir}
\end{align}
where $\rho_m(z)$ and $\rho_c(z)$ are the mean matter density and critical density 
at redshift $z$.  
For example, $\Delta_{vir} \approx 119$ at $z = 0.2$
($\Delta_{vir} \approx 273$ with respect to mean matter density).  
We list estimates of  $r_{vir}$ for individual galaxies in Table~\ref{tb:mass}.
The median halo virial radius for the 15 galaxies is 160~kpc, and the quasar
sightlines intersect the CGM at impact parameters of 0.1--0.5 $r_{vir}$.

The line-of-sight velocity dispersion through a halo of specified $M_{vir}$ and $r_{vir}$
depends on the mass distribution. 
In this section, for simplicity, 
we describe the halo profile as a singular isothermal sphere  
truncated at the virial radius $r_{vir}$.  
The resulting halo circular velocity,  
\begin{equation}
   V_{c,vir} = \sqrt{G M_{vir}(r_{vir})/r_{vir}}, 
  \label {eqn:v_c_sis} 
   \end{equation}
stays constant with radius.  
In virial equilibrium, the 3D velocity dispersion of 
the gas clouds would be $\langle v^2 \rangle = 3 k T / (\mu m_H)$. We adopt the isotropic
velocity dispersion in one dimension,
\begin{equation}
		\sigma_{v}^2 = \frac{\langle v^2 \rangle}{3} = \frac{V_{c,vir}^2}{2}
	\label{eq:vir-width}
\end{equation}
as our estimate of the line-of-sight velocity dispersion along the quasar sightline.

\begin{figure}[htb] 	
	\includegraphics[width=1.0\linewidth]{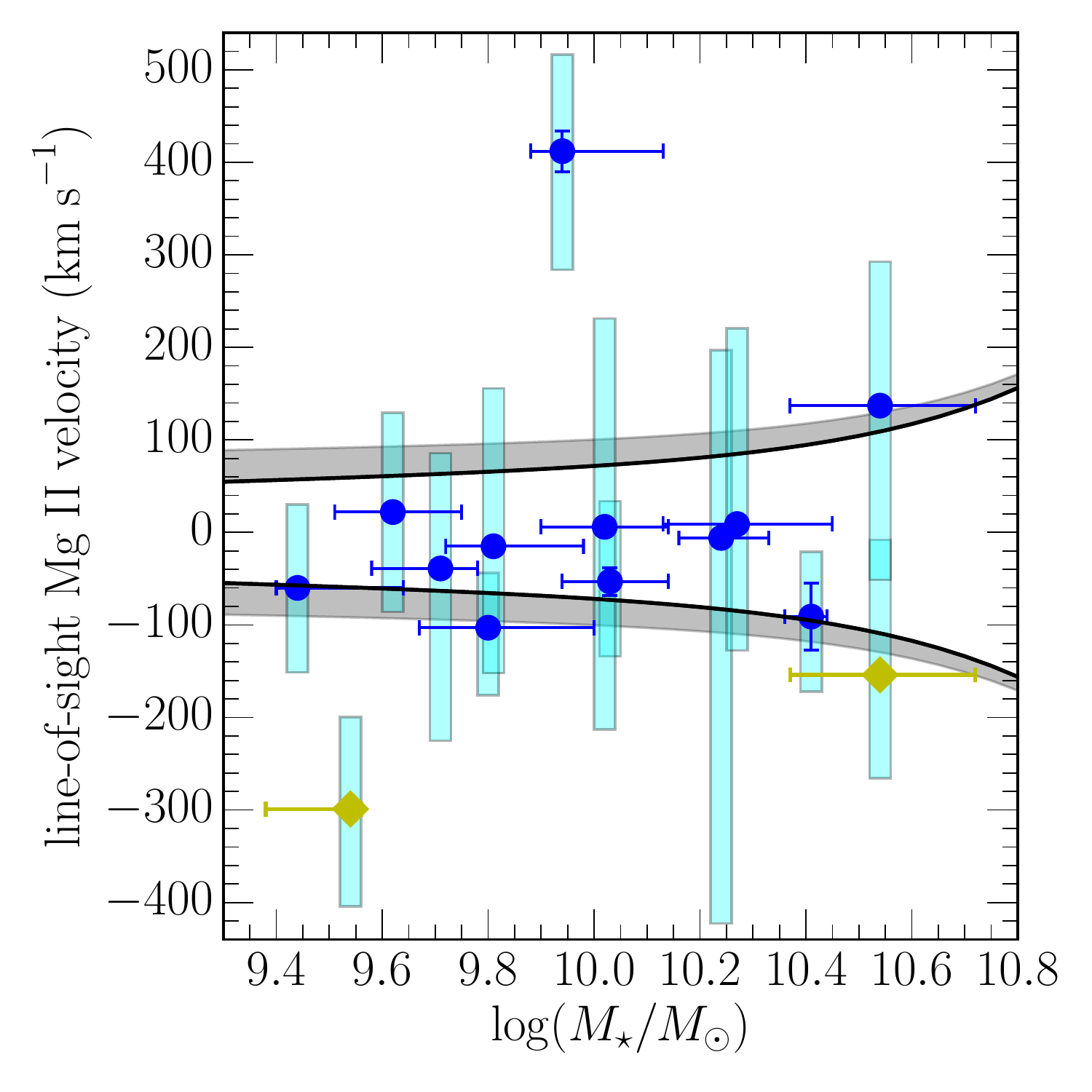}
	\caption{Velocity range of \mgII\ systems compared to halo velocity dispersion.
        The black lines illustrate the velocity dispersion of the halo (Equation~(\ref{eq:vir-width}))
		with an assumed redshift of $z = 0.2$; 
        the shading represents convolution with the spectral resolution.
        The observed velocity range of the \ion{Mg}{2}\ absorption 
        is consistent with the velocity dispersion of virialized clouds.
        Note that $\sigma_v$ is a mass-weighted mean velocity, so we expect
        some absorption at larger velocities.  All except one of the Doppler 
        velocities (blue points) lie within the expected range.
	}
        \label{fig:vvir_ms}
\end{figure}

Figure~\ref{fig:vvir_ms} illustrates the velocity dispersion of the \mgII\ systems over 
the mass range of our sample.  
\nedit{
Except for the \j103640+565125 sightline  
}
and the two systems already flagged as potentially associated with galaxies other than our primary target,
the velocity spread of the \mgII\ absorption along each of the other 11 sightlines 
is similar to the range predicted for clouds in virial equilibrium.  
In virial equilibrium, 
we would also expect the velocity range to broaden with increasing halo mass, and
the data do not contradict this expectation.

Our results probe a relatively small range in halo and galaxy properties, but
we find no contradictions to virial equilibrium in this regime.  In contrast,
absorber--galaxy cross-correlation studies have reported an inverse correlation between absorber strength 
(i.e., velocity width) and the  mean halo mass \citep{Bouche2006,Gauthier2009}, prompting much
discussion regarding the virialization of the clouds making up  \mgII\ systems 
\citep{Bouche2006,Chen2010,Tinker2010}. We note only that a more recent analysis of absorber--galaxy
catalogs shows there is no anti-correlation between \mgII\ equivalent width and virial mass
\citep{Churchill2013}.

\begin{figure}[htb]
	\includegraphics[width=1.0\linewidth]{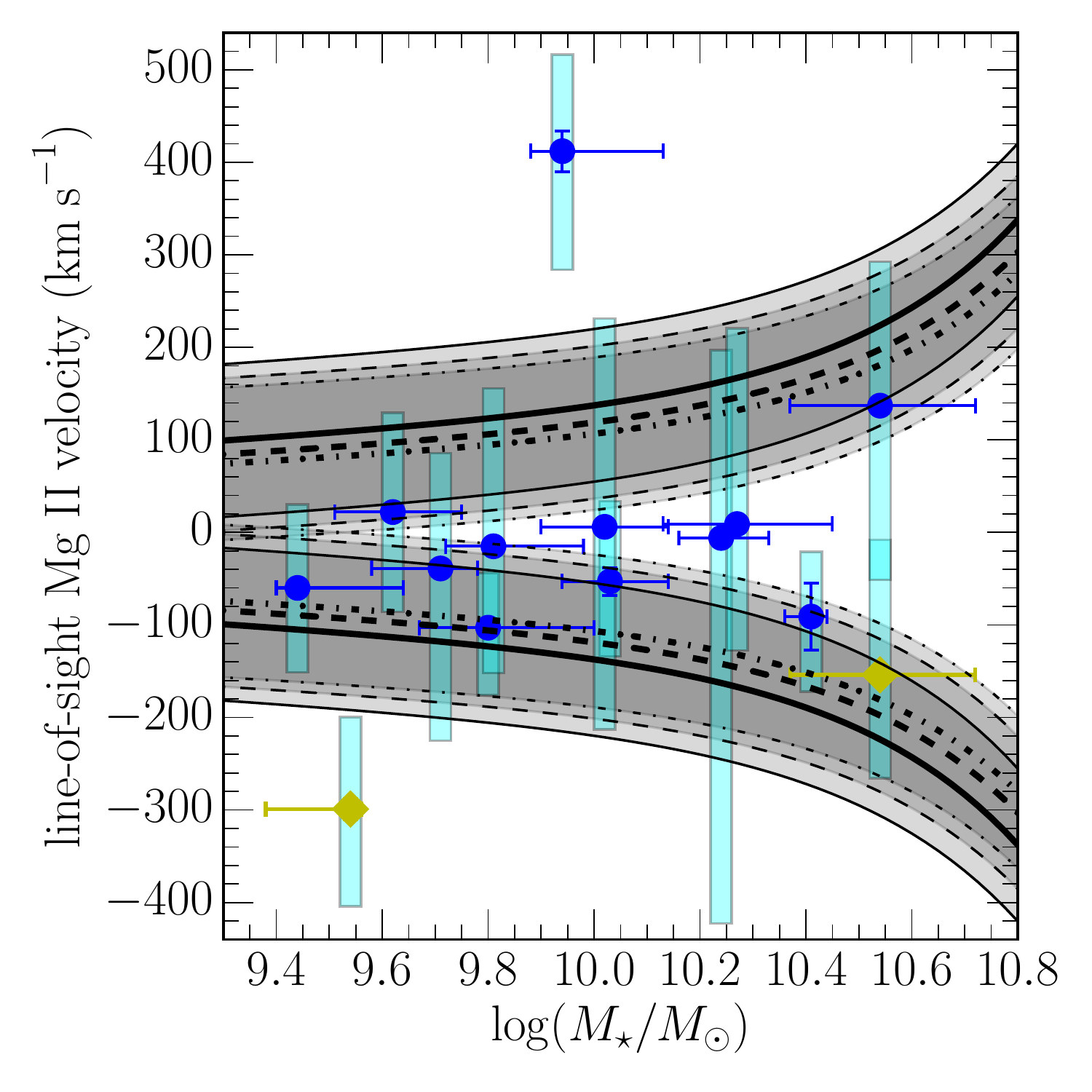}
	\caption{Comparison of low-ionization-state gas kinematics to the halo escape velocity.
          For purposes of illustration, we  assume  that the clouds move isotropically
          and plot the line-of-sight component of the escape velocity, $v_{esc}/\sqrt{3}$.
          A redshift of $z = 0.2$ is assumed for the models.
		  Symbols are as in Figure~\ref{fig:vvir_ms}.
		  The solid, dashed, and dotted--dashed lines are the escape velocites 
		  at galactocentric radii of 30, 60, and 90 kpc. 
		  The gray bands illustrate convolution with the line response function of the 
		  spectrograph.  For the pairs without large systemic velocity uncertainties,
		  most of the measured velocity ranges of the low-ionization-state gas are  
		  less than the escape velocity.
		  }
	\label{fig:vesc-ms} 
\end{figure}

Figure~\ref{fig:vesc-ms} compares the velocity width of the absorption troughs 
to the halo escape velocity, $v_{esc}(r) = \sqrt{2 |\Phi(r)|}$. The gravitational
potential of the halo, $\Phi(r)$, has been calculated for the isothermal halo profiles
at three radii chosen to span our range of the sightline impact parameter.  Clearly, 
the Doppler shifts of the  low-ionization-state absorption indicate that the clouds are expected to be 
bound to the halos. 
We add that directly matching the projected distance of each sightline with the escape velocity curves 
is a conservative comparison because the larger galactocentric radius of the cloud in 3D
places it in a region of the halo with lower escape velocity, $v_{esc}(r_{3D}) < v_{esc}(b)$.

\subsection{The Covering Fraction of Low-ionization-state Gas}

\begin{figure}[htb]
	\centering
	\includegraphics[width=0.85\linewidth]{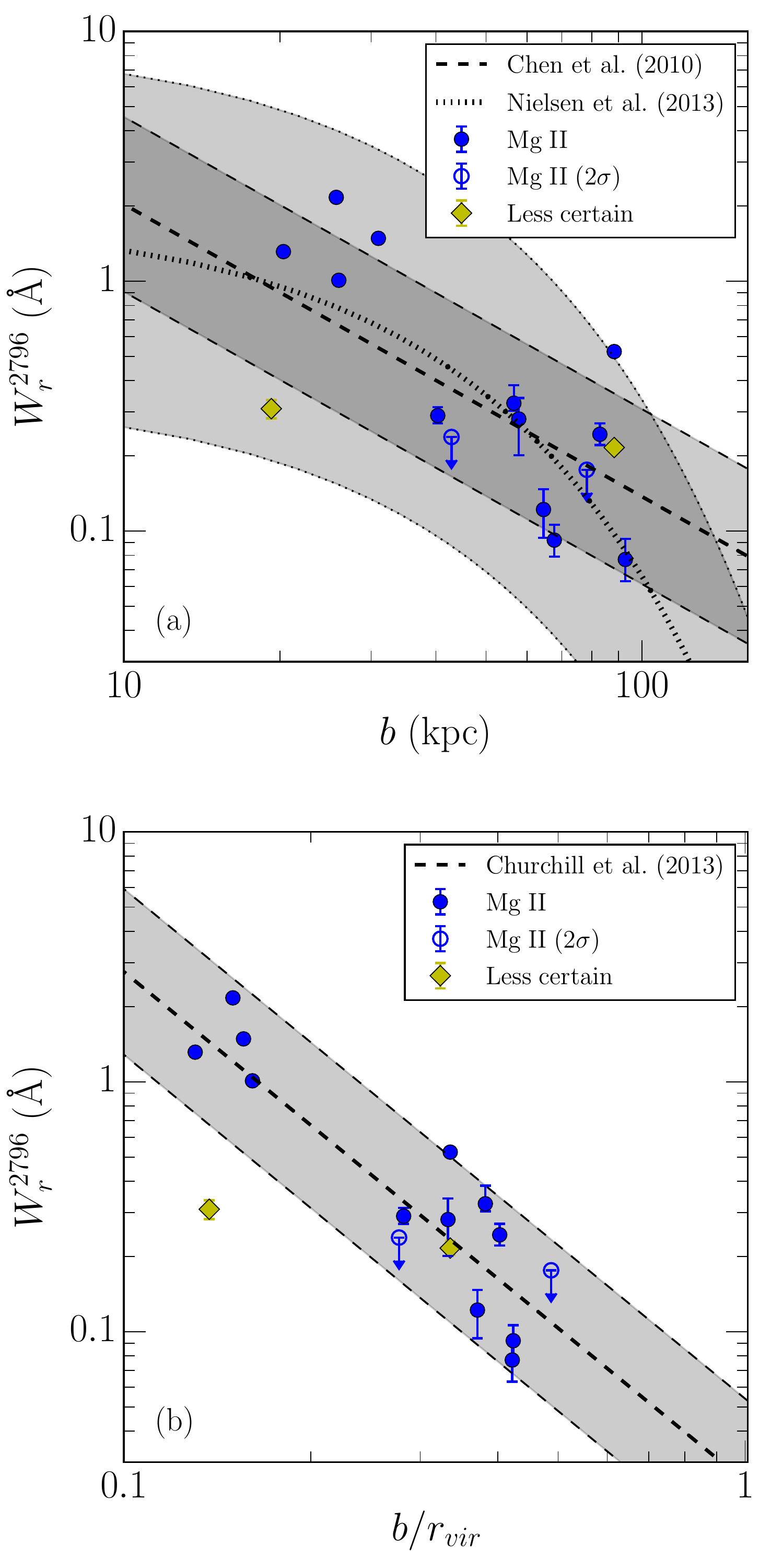}
	\caption{Decline in rest-frame equivalent width of \mgII\ absorption with 
        \nedit{(normalized)} impact parameter. 
		The filled markers represent detections, and 
		the unfilled markers show the $2\sigma$ upper limits for non-detections.  
		The yellow diamonds flag the uncertain systems as described in the text.  
		\nedit{In panel (a), }
        the thick dashed and dotted curves show the fitted correlations from
        \citet{Chen2010} and \citet{Nielsen2013}, respectively. The 
		shaded regions show the corresponding 
		root-mean-square residual between the data and the fit in \citet{Chen2010}, 
		and the  
		square root of the variance of the data relative to the fit in \citet{Nielsen2013}.  
		\nedit{
		In panel (b), the thick dashed line and the shaded region represent 
		the fitted relation and square root of the variance of data relative to the fit 
		in \citet{Churchill2013}.}
		}
	\label{fig:eqw_b} 
\end{figure}

\clm{
Of the 15 sightlines, seven show strong absorption, $W_r(\lambda 2796) \geq 0.3$~\AA\ 
\citep{Nestor2005}. Since all of our sightlines would easily detect strong absorption,
one could claim a covering fraction close to 50\%. However, the absorption strengths
decrease with impact parameter, following a well-known trend illustrated in
Figure~\ref{fig:eqw_b}(a).
Our covering fraction measurement accounts for both this radial trend and size
differences among the host galaxies.  In Figure~\ref{fig:eqw_b}(b), we normalize
the impact parameter by the halo virial radius. The strength of the detected
absorption is typical for the impact parameters. 
}

\clm{
Figure~\ref{fig:eqw_b} shows upper limits for the two non-detections. Since
these upper limits fall  within a standard deviation of the mean relation,
we cannot exclude typical \mgII\ absorption along these sightlines. Our
results are therefore consistent with a unity covering fraction. This high
major-axis covering fraction applies to the inner CGM. Our largest impact
parameters are 83~kpc (0.24~\AA, \j142501+382100) and 93~kpc (0.08~\AA, 
\j123049+071036).
}


\section{Discussion: Circumgalactic Gas Dynamics}
\label{sec:cgm}

We established in Section~\ref{sec:results} that our \mgII\ detections
probe circumgalactic gas largely in  virial equilibrium with the target galaxies.
However, the orbits of these gas clouds are not random, rather their Doppler shifts 
show a positive correlation with the sign of each galactic rotation curve.

These kinematic properties of the CGM appear to be consistent
with several decades of published literature on intervening absorption 
systems. For example, surveys of quasar fields traced intervening \mgII\ absorption to 
the halos of bright ($0.7L^*_B$) galaxies at intermediate redshifts, e.g., \cite{Steidel1994}. 
However, a deeper understanding of profile widths and their
notable asymmetries  \citep{Lanzetta1992} generated debate between proponents
of extended disk models \citep{Prochaska1997} and those favoring  more spherical 
halos \citep{McDonald1999}. As it turned out, adding a rotational component to the 
population of clouds proved to be the key to describing the statistical properties of
the \mgII\ line profiles. This solution, however, did not distinguish between rotating disks, 
halos, or hybrid descriptions \citep{Charlton1998}.

Our selection criteria --  specifically (1) inclined disks, (2) low azimuthal 
angles for the sightlines, and (3) low impact parameters -- clearly favor the 
detection of extended disks. Thus, rather than re-visiting the disk vs. spheroid
quandary with a small number of sightlines, we simply explore the conjecture that 
major-axis sightlines primarily select gas clouds in the extended plane of galactic disks. 
In this section, we further explore the relationship between the angular momentum 
of circumgalactic gas and galactic disks.

Our discussion focuses on the measured velocity range and centroid of \ion{Mg}{2} absorption 
in 11 CGM sightlines. We assume that \j091954+291345
produces the stronger of the two components in the \j091954+291408 
sightline and predict that one of the neighboring galaxies in the field will turn 
out to have a redshift closer to that of the weaker absorption component.
We  exclude the \nedit{flagged} \j124601+173156 sightline 
from the discussion in this section because
the \mgII\ absorption appears to be associated with a galaxy at smaller
angular separation from the quasar than our target galaxy.  
\nedit{We defer discussion of the \j103640+565125 sightline} until
we obtain high-resolution imaging because the other targets with such
high velocity components have turned out to be  groups or mergers.
We also note evidence
for a galaxy interaction with \j160951+353838, which may be related to 
the unusually broad and strong absorption.\\

\subsection{Angular Momentum of Circumgalactic Gas}
\label{sec:rotation}

\begin{figure*}[bth]
	\centering
	\includegraphics[width=0.9\linewidth]{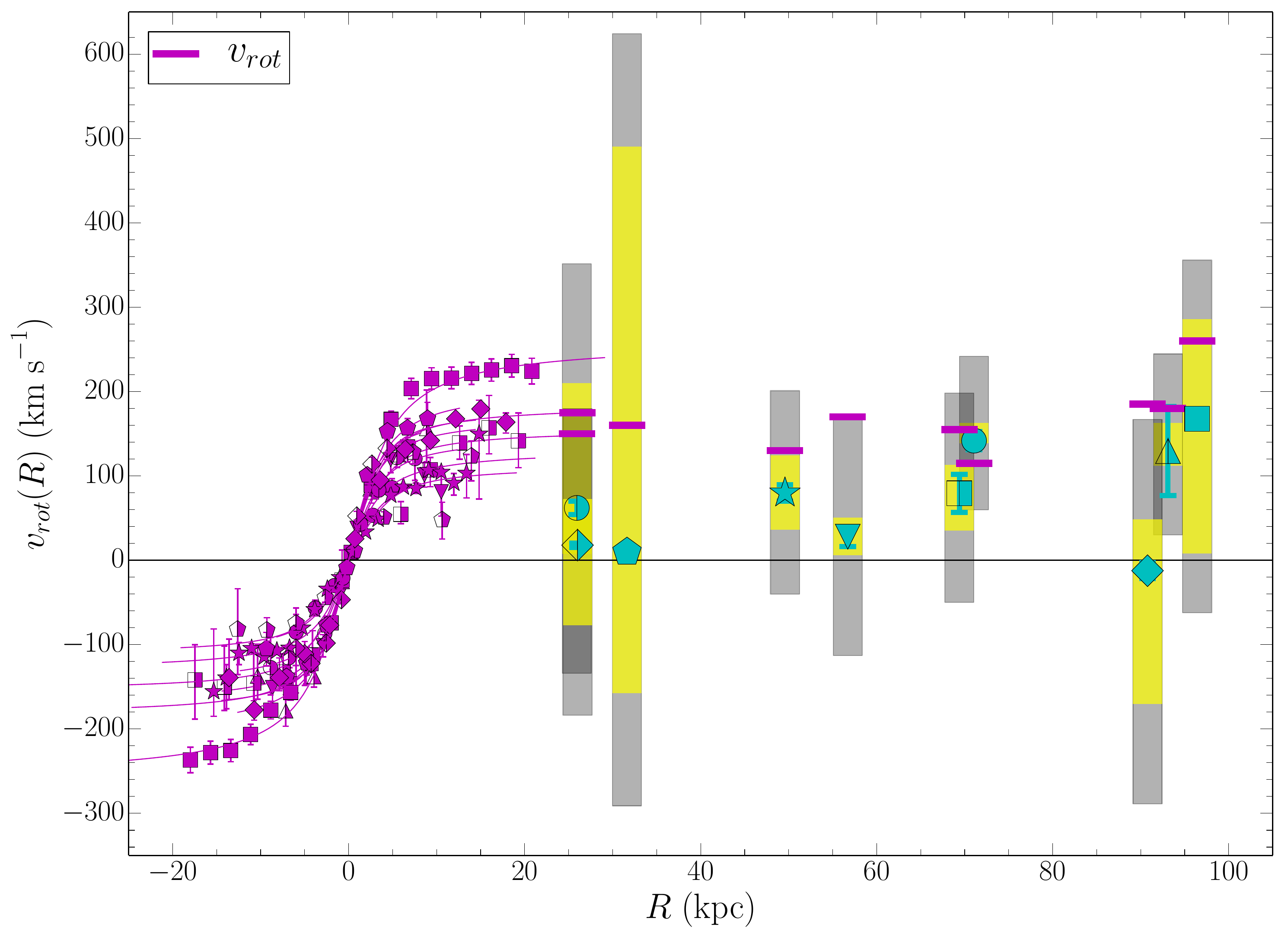}
	\caption{
		  Comparison of circumgalactic gas kinematics to galaxy rotation.  
		  For purposes of illustration, 
		  we have deprojected the Doppler shift and impact parameter of each \mgII\ system to a
          circular velocity and galactocentric radius. In other words, the \mgII\ velocity here 
          represents the tangential motion in the disk plane that would give the observed 
          Doppler shift when projected onto the quasar sightline. For purposes of uniformity in this
          figure, galactic rotation curves are defined with a positive velocity
          on the quasar side of the major axis. 
		  The symbol for each sightline is the same as in the PV-diagrams in Figure~\ref{fig:sdss_pv}, 
		  but all measurements are now deprojected onto the disk plane.   
		  The yellow (gray) bars indicate the deconvolved (measured) width of each absorption complex;
          these can be compared to the extended rotation curves indicated by the magenta bars.
		  Note that we omit \j160951+353843 in this figure and Figure~\ref{fig:peng_norm}
		  because the large absorption strength and width stand out from all the other sightlines; 
		  the measured velocity range requires expanding
		  the $y$-axis making it less possible to distinguish the individual rotation curves.
		  }
	\label{fig:peng}
\end{figure*}

\begin{figure*}[bth]
	\centering
	\includegraphics[width=0.9\linewidth]{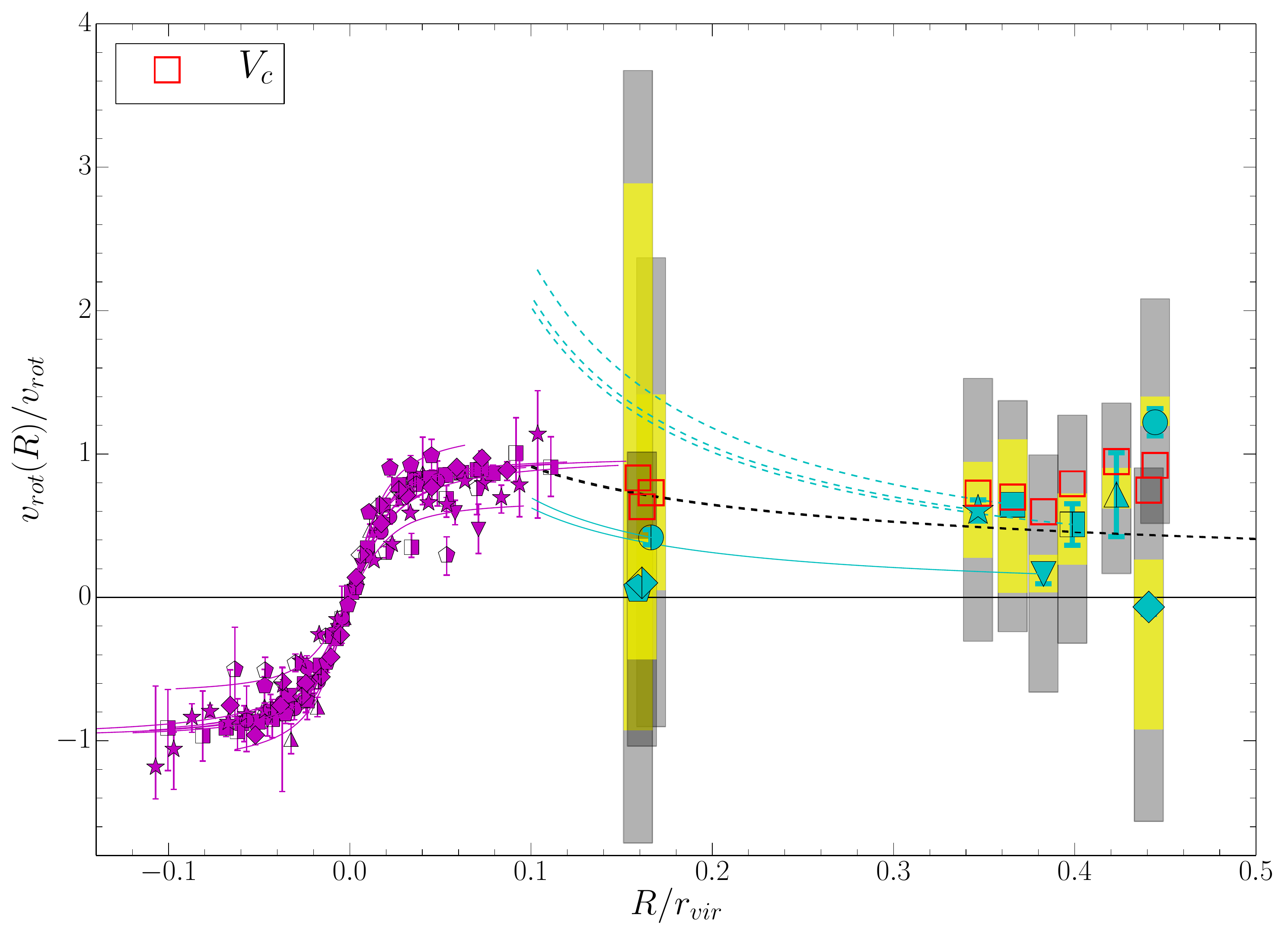}
	\caption{Comparison of disk and halo angular momentum.  
          Similar to Figure~\ref{fig:peng}, but this diagram shows the 
		  normalized rotation speed as a fraction of the halo virial radius. The points
          show our models of the actual rotation speed in the halo accounting for the dark matter 
          halo (red squares); the Keplerian fall off from the measured rotation curve sets a lower
          limits on the rotation speed in the CGM (dashed, black line).
          The cyan curves illustrate constant $Rv_{rot}(R)$ and show that the infalling
          gas would have specific angular momentum at least as large as that in the galactic disk.  
		  }
	\label{fig:peng_norm}
\end{figure*}

Our observations do not directly determine the location of the clouds along each sightline. 
To gain insight about their relationship to the galaxy, we  
\clm{\nedit{consider}}
the conjecture that the clouds populate the extended 
plane of the galactic disk, calculate their implied galactocentric radius and circular velocity, 
and then examine the implications for the dynamical state of the gas.

Figure~\ref{fig:peng} compares the implied rotation speed of the \mgII\ systems to the 
rotation speed of the galactic disk. As described in Section~\ref{sec:rotation}, the
Doppler shifts of the \mgII\ systems share the sign of the galactic rotation along seven
sightlines. The measurement errors for four systems are consistent with no net
Doppler shift, and  these systems have substantial equivalent width
on both sides of the systemic velocity. The range of deprojected \mgII\ velocities often reaches
the asymptotic rotation speed of the disk. The Doppler shift of most of the
absorption equivalent width is, however, too low to be consistent with purely circular orbits in a disk.

In Figure~\ref{fig:peng_norm}, we have normalized the galaxy rotation curves 
by the asymptotic rotation speed, $v_{rot}$, of each galaxy.  
The red squares show the halo circular velocity assuming 
a Navarro, Frenk, and White (NFW; \citealt{Navarro1996}) halo profile 
with the concentration parameter $c(z,M_{vir})$ calculated using the
python package \texttt{Colossus} introduced by \cite{Diemer2015}.\footnote{
        \citeauthor{Diemer2015}~(\citeyear{Diemer2015}, hereafter DK15) adopt $M_{200c}$ 
		to describe a halo with a mean density 200 times 
        the critical density and model the corresponding concentration parameter $c_{200c}$, 
		but \cite{Behroozi2010} define halo mass differently.  
		\texttt{Colossus} provides the conversion of halo concentration between different 
		mass definitions, for which
		the discussion in Appendix C of DK15 suggests that the inaccuracies of these conversions 
        are no more than  $\sim$20\%.
		}
The halo circular velocity resembles the form of Equation~(\ref{eqn:v_c_sis}) but with a radial dependence,
\begin{equation}
   V_c(r) = \sqrt{G M_{vir}(r)/r}. 
  \label {eqn:v_c} 
   \end{equation}
Most of the \mgII\ equivalent
width is detected at Doppler shifts less than that generated by clouds on circular orbits.
The velocity widths of the systems are broad enough to include gas on a circular orbit, but 
the estimated speed of a circular orbit exceeds the centroid velocity 
(cyan symbols in Figure~\ref{fig:peng_norm}).
Adding the stellar mass from Table~\ref{tb:mass} to the halo mass raises the circular velocity slightly
(2 - 6\%) and increases the magnitude of the discrepancy.
Only the sightline probing the CGM at the largest radius provides an exception. At
$R/r_{vir} \approx 0.45$, the \j084723+254105 sightline has a Doppler shift larger
than expected from circular orbits.

If clouds near the disk plane with $R/r_{vir} \sles 0.45$ are not fully described by
circular orbits, then what can we say about their dynamical state? Since we have shown
that the velocity spread of the systems is consistent with virial motion, we can
conclude that the clouds have a significant velocity component that is perpendicular
to any tangential motion in the disk plane. We suggest radial inflow as a physically
likely origin for this extra velocity component. While the spectral line profiles
do not uniquely distinguish radial inflow from other velocity vectors, we argue
that our selection of major-axis sightlines favors this solution.

If our interpretation is correct, then it has implications for how galaxies get their
gas. We illustrate where the infalling gas might obtain a circular orbit by drawing
curves of constant angular momentum in Figure~\ref{fig:peng_norm}. By following these
curves, we see that some of the gas detected in \mgII\ absorption has specific 
angular momentum comparable to the galactic disk. The broad line widths, however,
indicate that other clouds might obtain circular orbits at radii several times
that of the visible galactic disk.  Large gas disks have been detected in \ion{H}{1} 21 cm
observations of nearby spiral galaxies. If we have detected extended gas disks
at $z \approx 0.2$, then the composition of the gas is clearly not
pristine implying significant metal recycling.

\subsection{The Tension between the Velocity Widths of \mgII\ Absorbers and Rotating Disks}
\label{ssec:compare-vw}

\begin{figure*}[htb]
	\centering
	\includegraphics[width=0.9\linewidth]{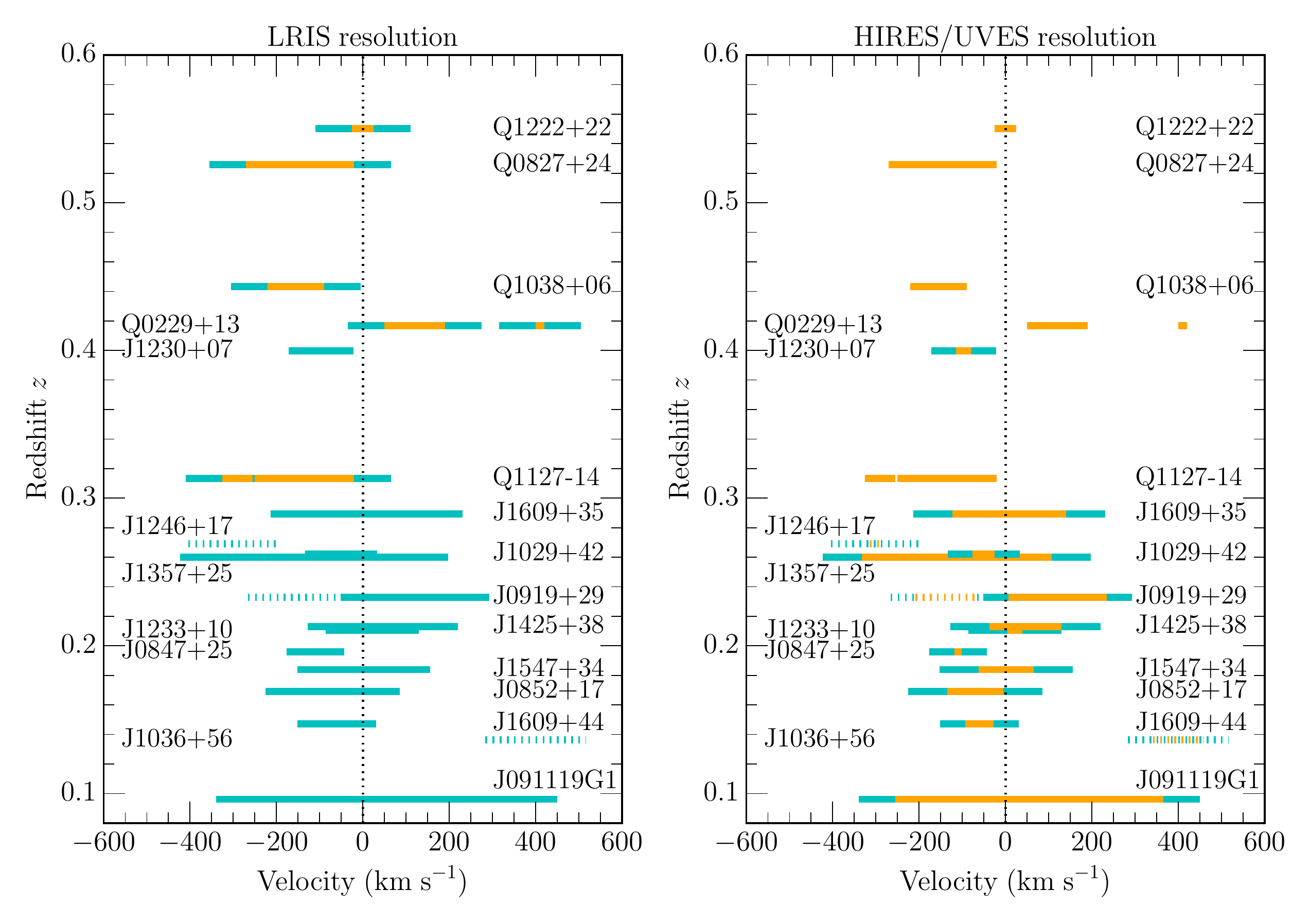}
	\caption{Velocity range of major axis \mgII\ absorption. Our results for
        11 sightlines (solid bars) are compared to previously published measurements over a 
        range of redshift. Significant absorption is detected to both sides
        of the systemic velocity for all low redshift galaxies observed; in contrast,
        the velocity ranges are one-sided in S02 and K10.
        {\it Left Panel:} Comparison at moderate spectral resolution. 
        Broadening of the echelle spectra by the LRIS instrumental
        resolution is illustrated by the cyan bars.  
        {\it Right Panel:} Comparison at high spectral resolution.
        To illustrate the LRIS instrumental broadening (cyan), we
        scale the correction factor derived in K11 to the LRIS spectral
        resolution of each run. The orange bars are our best estimate of
        the intrinsic system velocity spread.
		Dashed lines represent the excluded absorption components or sightlines 
		in our analysis in Section~\ref{sec:cgm}.
		}
	\label{fig:vw} 
\end{figure*}

Figure~\ref{fig:vw} illustrates the velocity spread of our \mgII\ absorption troughs
relative to the systemic velocities of the associated galaxies.  We have added
published observations of sightlines with $\alpha \le 30\deg$ near inclined, star-forming galaxies.
We do not include sightlines from the COS Halos survey \citep{Tumlinson2011,Tumlinson2013, Werk2013,Werk2014} because
no measurements of galaxy rotation have been published.  
As introduced in Section~\ref{sec:rotation}, three sightlines from 
\citeauthor{Steidel2002}~(\citeyear{Steidel2002}, hereafter S02) satisfy our selection criteria.
At intermediate redshifts, we add two sightlines from 
\citeauthor{Kacprzak2010}~(\citeyear{Kacprzak2010}, hereafter K10); and, 
at lower redshift, we found one sightline  from \citeauthor{Kacprzak2011ApJ}~(\citeyear{Kacprzak2011ApJ}, hereafter K11).  
Little data of this type exists at higher redshifts, 
but two ground-breaking papers suggest that similar trends may be present in the low-ionization-state absorption.  
Just 20\deg\ off the major axis of a $z = 2.3283$ galaxy, 
\cite{Bouche2013} detected absorption from low ions (but did not cover \mgII) at an impact
parameter of 26~kpc. 
The Doppler shift of the main absorption component is 180 \kms\  
and in the same direction as the galactic rotation, whereas the 
total absorption spans from $-35$\kms\  to $+300 $\kms.
Just 12~kpc from a $z = 0.9096$ galaxy,
\cite{Bouche2016} resolve the \mgII\ line profile in a sightline 15\deg\ off the major axis. The Doppler shift
is again in the same sense as the galaxy rotation, and the velocities range from $-230$\kms\ to $+265$\kms.

Differences in instrumental resolution among these observations affect the measured velocity widths
of the absorption troughs. We have attempted to correct all observations to the same effective
resolution in Figure~\ref{fig:vw}.  In the left panel of Figure~\ref{fig:vw}, we have simply degraded the spectra
of bright quasars taken using Keck/HIRES or VLT/UVES ($v_{\mathrm{res}} \simeq 6$ \kms) to
our resolution by convolving with the LRIS line-spread function.  
The inverse process --  recovering the intrinsic velocity width of a system from our lower resolution
spectra -- is inherently more noisy. We show the results of this exercise in the right panel of
Figure~\ref{fig:vw}.  The observations reported by K11 were also obtained with Keck/LRIS at   
a spectral resolution ($v_{\mathrm{res}} \simeq 155$ \kms) similar to our median resolution. 
Section 4.1 of the K11 paper convincingly argues that these LRIS line widths should be decreased 
by 85 \kms\ on both sides before they are compared to well resolved line profiles. We have confirmed
that simply fitting the \mgII\ absorption with a Gaussian profile convolved with
the LRIS line spread gives a 
\clm{
consistent velocity width.}  After taking instrumental broadening
into account, we find that four of our 11 systems  show absorptions at velocities on 
both sides of the galaxy systemic velocity.

A dynamical description of the low-ionization-state gas kinematics must explain the following:
(1) the correlation of the sign of the net Doppler shift with the galactic disk, (2) the broad
line width, and (3) the generation of absorption on both sides of the systemic velocity. The
simple disk model naturally satisfies the first criterion, but the other kinematic properties
prove challenging to explain with the simple disk picture. For individual systems, we have 
explored two types of solutions: (1) adding a vertical velocity gradient to the disk rotation 
speed and (2) adding radial infall in the disk plane.  We provide a mathematical description 
of these models in the Appendix. The solution space is highly degenerate. Rather than presenting 
every possible solution, we focus on answering a few well posed questions.

\subsubsection{Are There Sightlines that cannot be Fitted by Disk Rotation?}

Disk rotation can never produce a sign change in the line-of-sight velocity along a sightline.
Equation~(\ref{eqn:v_los_sk}) and diagrams in the Appendix
demonstrate this fact quantitatively.\footnote{ 
     Poor spectral resolution could certainly smear the line profile  so
     that it crossed the systemic velocity. In principle, turbulence could provide a physical
     mechanism to accomplish this line broadening, but requiring turbulent velocities 
     comparable to the circular velocity  implies a structure that cannot really
     be labeled a disk.}

We find four sightlines with absorption on both sides of the systemic velocity -- 
\j135734+254204, \j142501+382100, \j154741+343357, and \j160951+353843, 
\clm{
so}
the simple disk model cannot describe the velocity range of the \mgII\ absorption.
The spherical halo model adequately describes these systems, and these sightlines need not
intersect a disk at all. Nonetheless, we asked whether an extended disk with radial
inflow provided a viable alternative explanation of the line widths.  
\nedit{
We found that inflow ($v_R = -250$\kms) in  a thick  disk ($2 H_{\mathrm{eff}} = 40$~kpc) 
produces the velocity range of the \j142501+382100 system.
The other three systems required thicker disks and higher inflow speeds. The
extreme values of these parameters lead us to conclude that extended disks
are not a viable model for these systems.
}

For the systems with single-sided absorption, a rotating disk can describe the velocity 
range measured in five of the seven sightlines, albeit with a troubling implication.
Recall that in Figure~\ref{fig:vw}, the instrinsic line widths typically span 
around a hundred \kms\ or more.  
The problem is that the pathlength of a sightline through a thin disk is just the vertical thickness 
of the disk, $2 H_{\mathrm{eff}}$ (as illustrated in the schematic diagrams in the Appendix), 
lengthened by the secant of the viewing angle. Unless
we view the disk exactly edge-on, the intercepted velocity range will be small.  
we found that the simple disk model required disk thicknesses of the order of the virial
radius -- i.e., tall rotating cylinders rather than disks.

As demonstrated in S02 and in the Appendix, 
adding a scale-height $h_v$ for the vertical velocity gradient 
such that the disk rotation speed decreases with increasing perpendicular distance from 
the disk midplane
will bring the line-of-sight velocity $v_{los}$ toward zero 
over a shorter pathlength along the sightline. 
As a fiducial reference point, we adopted $h_v = 10$~kpc.  
For our measured range of rotation speeds, this choice of 
velocity scale height produces a vertical gradient between 11 and 26\kms\ per kpc.
\nedit{
Above the plane of nearby spiral galaxies, the rotation lags the disk; 
our adopted $h_v$ creates a vertical velocity gradient consistent 
with measurements of $dv_{\phi} / dz = -8 {\rm ~to~} -60$ \kms\ per kpc
\citep{Oosterloo2007,Heald2011,Marasco2011,Zschaechner2011,Zschaechner2012,
Gentile2013,Kamphuis2013,Zschaechner2015}.  
}
With this velocity gradient in
the extraplanar gas, we found solutions with very thick disks and
reduced the median $H_{\mathrm{eff}}$ to 20~kpc 
(for the models fitted to the five disks).\footnote{
	{The total vertical thickness of the disk is $2H_{\mathrm{eff}}$; 
	$H_{\mathrm{eff}}$ measures the disk thickness from the disk midplane.}
	} 
We also found an alternative picture, however, which appears at least as plausible
in our opinion. When a radial inflow component was allowed, the median $H_{\mathrm{eff}}$
dropped to 10~kpc, and the range of inflow speed ranged from 40 to 180\kms.

\begin{figure*}
		\centering
		\includegraphics[width=1\linewidth]{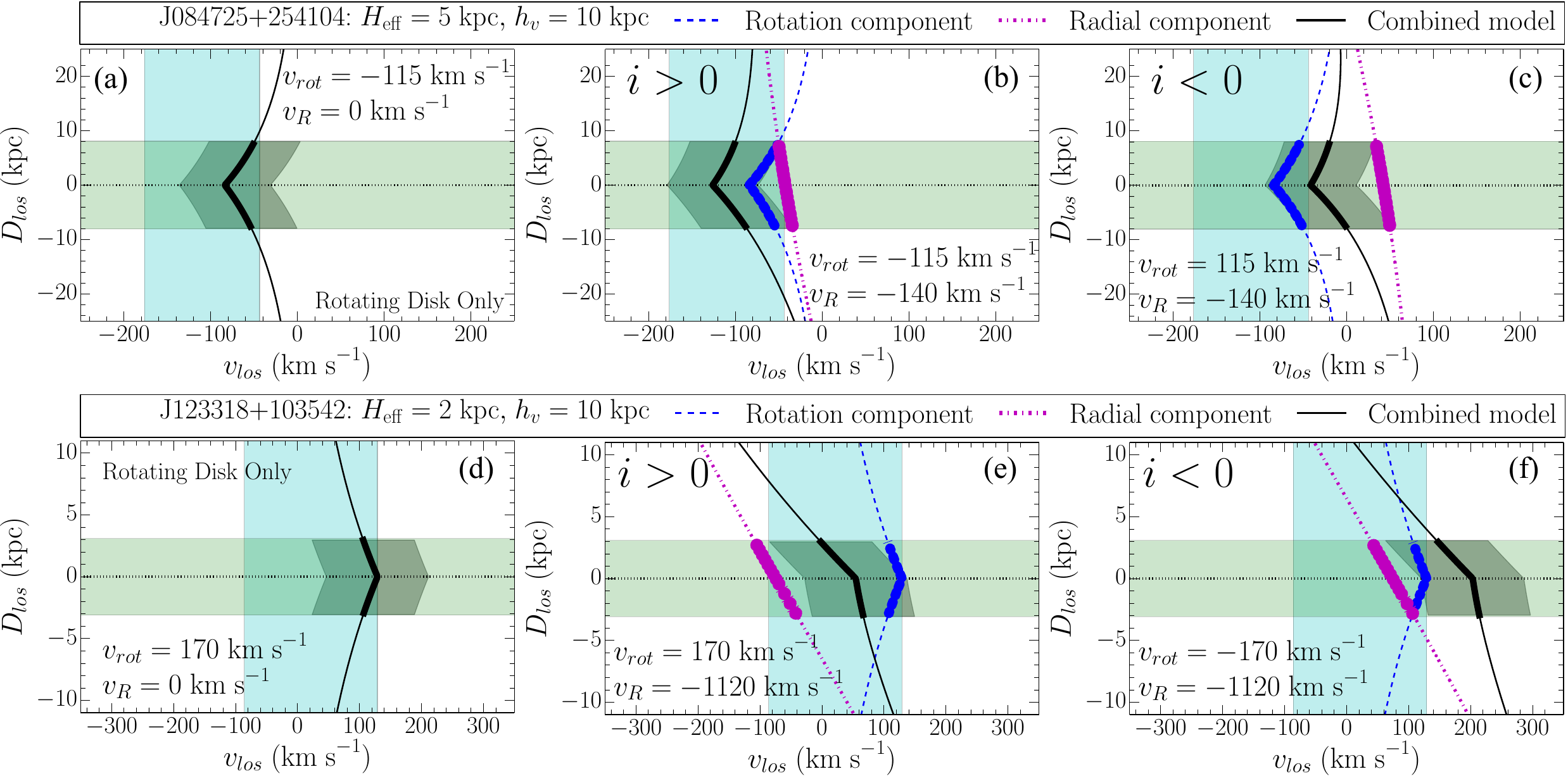}
                \hspace{10cm}
		\caption{Examples of the infall solution for the kinematics of 
                  absorption-line systems.
                  The cyan shading shows the observed line-of-sight velocity range for 
                  galaxy \j084725+254104 (top row) and \j123318+103542 (bottom row) 
				  along their corresponding quasar sightlines. 
                  Left columns illustrate the failure of the simple disk model.
                  The disk of \j084725+254104 rotates too slowly to produce the large
                  Doppler shifts observed. The rotation speed of the \j123318+103542
                  disk creates a large Doppler shift than we observe.
                  Middle and right columns contrast the two signs of the disk inclination.
                  To reproduce the velocity range of the \mgII\ absorption (cyan), 
                  the inflow model (bold black line) requires a positive sign of the disk inclination.
                  The opposite tilt of the disk produces absorption that is too redshifted 
                  along both sightlines.  
				  We adopt the right-handed coordinate system for which $v_{rot}>0$ 
				  produces an angular momentum vector along the positive $z$-axis of the disk, where the 
				  positive $z$-axis always points toward the quasar side of the disk 
				  and is perpendicular to the disk plane.   
				  The sign flip of the disk inclination therefore changes the sign of $v_{rot}$ in the model   
				  (see the Appendix for more detail).  
				  In each position--velocity diagram ($D_{los}$ vs. $v_{los}$),
				  the green shaded region indicates the portion of the sightline 
				  intersecting the disk.  The gray band illustrates the convolution of the modeled
				  line-of-sight velocity with the line response function of the spectrograph.  \\
			}
		\label{fig:model1}
	\end{figure*}

Figure~\ref{fig:model1} illustrates the failure of the thick disk model for
the other two \mgII\ detections. The thick disk model cannot describe
the \j084723+254105 system because the large blueshift of the absorption 
exceeds the projected rotation speed ($v_{rot} \sin i = 115\kms\ \sin (52\deg) = 91\kms $).
The bottom row of Figure~\ref{fig:model1} illustrates  a slightly
different mode of failure for the thick disk model.  
Toward the sightline associated with \j123318+103542, 
the absorption velocities do not reach the projected circular velocity of the galactic disk. 
In the center panels of Figure~\ref{fig:model1}, we successfully fit these two
systems by introducing radial infall in the disk plane. The observed line-of-sight 
velocities are the projection of the total velocity vector produced by the addition of
rotation and inflow in the disk plane. Whether the infall boosts or cancels the 
projected circular velocity depends on geometry. 
The Appendix clearly illustrates 
the origin of this asymmetry; we could not find it described 
previously in the literature. 
This solution has a reasonable disk of 
thickness $H_{\mathrm{eff}} = 5$~kpc and $H_{\mathrm{eff}} = 2$~kpc, 
with the velocity scale height $h_v = 10 $~kpc in the models 
for \j084725+254104  and  \j123318+103542 respectively.  
The corresponding inflow speeds are $v_R = -140$\kms\ and $v_R = -1120$\kms\ in the two models.

\subsubsection{Is Radial Inflow Excluded When it is not Required?}

A rotating disk with radial infall provides a plausible description of the 
\mgII\ kinematics in our two systems that cannot be fit with simpler models.
While the other systems do allow simpler descriptions, we emphasize that
the implied parameters generate tension with our expectations for real disks.
Without inflow, the five systems described by rotating disks require very 
thick disks, and we suggest solutions with thinner disks and inflow as a viable alternative.
The same picture, inflow in a rotating gas disk, can reproduce broad absorption
centered near the systemic velocity. We do not favor this interpretation for those four 
sightlines because simpler models remain consistent with our observations; we
simply emphasize that, {\it no}, we do not exclude this interpretation
for a single sightline.


\subsection{Implications for How Galactic Disks Get Their Gas}

The gas flows in simulations of galaxy formation motivated us to consider the observational 
signature of gas flows near the plane of the galactic disk. We observed the CGM along sightlines
near the galactic major axis in an effort to select such disks if they exist.  
We demonstrated that this circumgalactic gas 
has a component of angular momentum in the same direction as the disk angular momentum.
If the gas clouds follow circular orbits in the extended plane of the galactic disk, however,
their Doppler shifts would be larger than observed. One interpretation of this result 
is simply that the angular momentum vector of the inner circumgalactic gas is only partially
aligned with that of the disk.  We suggest, however, that radial inflow in the disk plane 
provides another viable description of the observations.

\clm{
The inferred mass flux depends directly on the column
density. Since we generally obtained a very conservative lower limit
on the column density, we estimate a lower limit of the mass flux. 
For purposes of illustration,
we use solar metallicity \citep{Morton2003} and 
a unity ionization fraction of singly ionized magnesium, 
i.e., $N(\mathrm{Mg}^{+})/N(\mathrm{Mg}) = 1$.\footnote{
	\nedit{The ionization fraction $N(\mathrm{Mg}^{+})/N(\mathrm{Mg})$ 
	varies between 0.1 and 1\citep{Murray2007,CLMartin2012}.}
	}
Substituting the lower limits from Table~\ref{tab:lris_result}, 
we infer hydrogen columns from $6\times10^{16}$\col\ to $3\times10^{18}$\col.
We apply Equation~(1) from \cite{Bouche2013}. For the two systems that require
inflow, the mass flux exceeds 0.07--1 \msunyr. If we interpret  
the five systems with single-sided absorption as inflow detections,
then the mass fluxes are at least 0.02--0.08 \msunyr. 
\nedit{In general,}
the true mass flux could be much, much larger; but we will need 
higher resolution spectra to produce accurate measurements.
}

In the context of this inflow picture, the new observations place some constraints
on how galaxies get their gas. First, the inflow speeds we found were generally
within a factor of two of the rotation speed. One sightline, 
quasar \j123317+103538 associated with galaxy \j123318+103542, required
a much larger inflow velocity to cancel the tangential velocity. We remain skeptical
of inflow velocities as large as 1000\kms; however, some high velocity clouds have velocities 
this large relative to the Local Standard of Rest \citep{Wakker1997}.  
Second, in Figure~\ref{fig:peng_norm}, we 
sketched curves of constant $R V(R)$ and then argued that some of the infalling gas would 
have to reach the galactic disk.  Possible strategies for testing this picture
include the following: (1) confirming/refuting the predictions
for the disk inclination and spiral arm morphology, (2) examining the implications
of the model for a broader range of azimuthal angles, and (3) significantly increasing
the sample of major-axis sightlines.

Given the absence of previous evidence for gas inflow in galactic disks, this result if
verified would have  significant implications for how galactic disks get their gas. In
this context, it is interesting to compare the galaxies in our sample to a younger version
of the MW galaxy in the past.
We use the stellar mass evolution function presented in \citet{vanDokkum2013}.
They study galaxies at different redshifts with the same rank order in stellar mass 
as the MW at $z = 0$, and adopt the stellar mass of the present-day MW as around 
$5\times10^{10}M_\odot$, i.e., $\log(M_\star/M_\odot) = 10.70$
\citep{Flynn2006,McMillan2011}.  
They then associate galaxies at different redshifts 
by requiring them to have the same cumulative comoving number density.  
We use their approximate stellar mass evolution function 
\begin{equation}
	\log({M_{\star\textrm{,MW}}/M_\odot}) = 10.7 - 0.045 z - 0.13 z^2
	\label{fig:mw-evolution}
\end{equation}
to find the predicted stellar mass of the MW progenitors at $z = 0.2$ 
as $\log({M_{\star\textrm{,MW}}/M_\odot})_{z=0.2} = 10.69$, which is marginally lower than 
that at $z=0$.  
This predicted stellar mass has an uncertainty of $\sim$0.2 dex.
The comparison between our galaxy sample at $z=0.2$ and 
the MW progenitors at $z = 0.2$ show that our median galaxy is 
\nedit{
0.87 dex
}
less massive
than the typical MW progenitor 2.5~Gyr ago; but the upper mass range of our sample is 
consistent with the expected masses of MW progenitors.

\section{Conclusion}
\label{sec:conclusions}

We presented new observations of 15 galaxy--quasar pairs. This study more than
doubles the number of quasar sightlines studied within $30^{\mathrm{o}}$ of the major axis
of star-forming galaxies. The focus on typical, star-forming galaxies allows us to consider
the ensemble of sightlines as multiple sightlines through the same average CGM.

Gas clouds in the plane of these galactic disks are
a plausible source of \mgII\ absorption detected in the quasar spectrum near the redshift 
of the foreground galaxy. Models predict that such gas has been accreted recently. We 
therefore asked the question of whether the gas kinematics might yield signs of accretion.

We detected \ion{Mg}{2}\ absorption in 13 sightlines 
\nedit{with $W_r(\lambda 2796) \gtrsim 0.1$ \AA.}
The sign of the Doppler shifts of these
systems always matched the sign of the galactic rotation on the quasar side of the major axis. This
result demonstrated that the motion of the absorbing gas is not random. The observations
do not require the absorbing clouds to be located in the plane of the galactic disk, but we 
focused our modeling effort on gas in an extended plane for two reasons: (1) these models
offer the simplest geometry allowed by the data, and (2) numerical simulations of individual
galaxies predict that gas accreted at late times feeds an extended gas disk.
\clm{
The Doppler shifts of the \mgII\ systems are less than expected for gas on circular orbits.
If these clouds reside in the disk plane, then they will spiral inwards through the disk.
This inflow mechanism would broaden the absorption troughs.
}

\nedit{
The velocity widths of the \mgII\ systems typically agree with expectations for
gas in virial equilibrium. Host galaxy mis-assignments plausibly explain the outliers.
However, the velocity range poses a challenge for simple models.}
The velocity widths of two \mgII\ systems cannot be fit
by a thick rotating disk model. Furthermore, modeling the widths of many systems requires
the disk to be ridiculously thick, essentially morphing into a rotating cylinder that is
much thicker than the disk radius.

We found that radial infall in the disk plane solves the above dilemmas. Other solutions may
exist, but we argue that radial infall is the simplest geometry consistent with the data.
These measurements provide a benchmark against which the accuracy of the gas accretion
process in numerical simulations should be evaluated. Our future work will address how
the circumgalactic gas kinematics depend on azimuthal angle.

\acknowledgements
This research was partially supported by the National Science Foundation under AST-1109288.  
G.G.K.~acknowledges the support of the Australian Research Council through 
the award of a Future Fellowship (FT140100933).  
\nedit{
We thank the referee for the useful suggestions.
}
We thank Mark Seibert for the discussion on GALEX photometry. We gratefully acknowledge
conversations with Peng Oh and Andrey Kravtsov, which  contributed to shaping our interpretation 
of these data. Most of the data presented herein were obtained at the W.M. Keck Observatory, 
which is operated as a scientific partnership among the California Institute of Technology, 
the University of California and the National Aeronautics and Space Administration. 
The Observatory was made possible by the generous financial support of the W.M. Keck Foundation.
The authors wish to recognize and acknowledge the very significant cultural role and 
reverence that the summit of Mauna Kea has always had within the indigenous Hawaiian community.  
We are most fortunate to have the opportunity to conduct observations from this mountain.  
Some presented data are
based on observations obtained with the Apache Point Observatory 3.5 m telescope, 
which is owned and operated by the Astrophysical Research Consortium.\\

{\it Facilities}: \facility{Keck:I (LRIS)}, \facility{Keck:II (NIRC2)}, \facility{Gemini:Gillett (GMOS)}, \facility{GALEX}, \facility{ARC}

\bibliography{master}

\appendix
\section{Appendix on Models}

\begin{figure}[bht]
	\centering
	\includegraphics[width=0.4\linewidth]{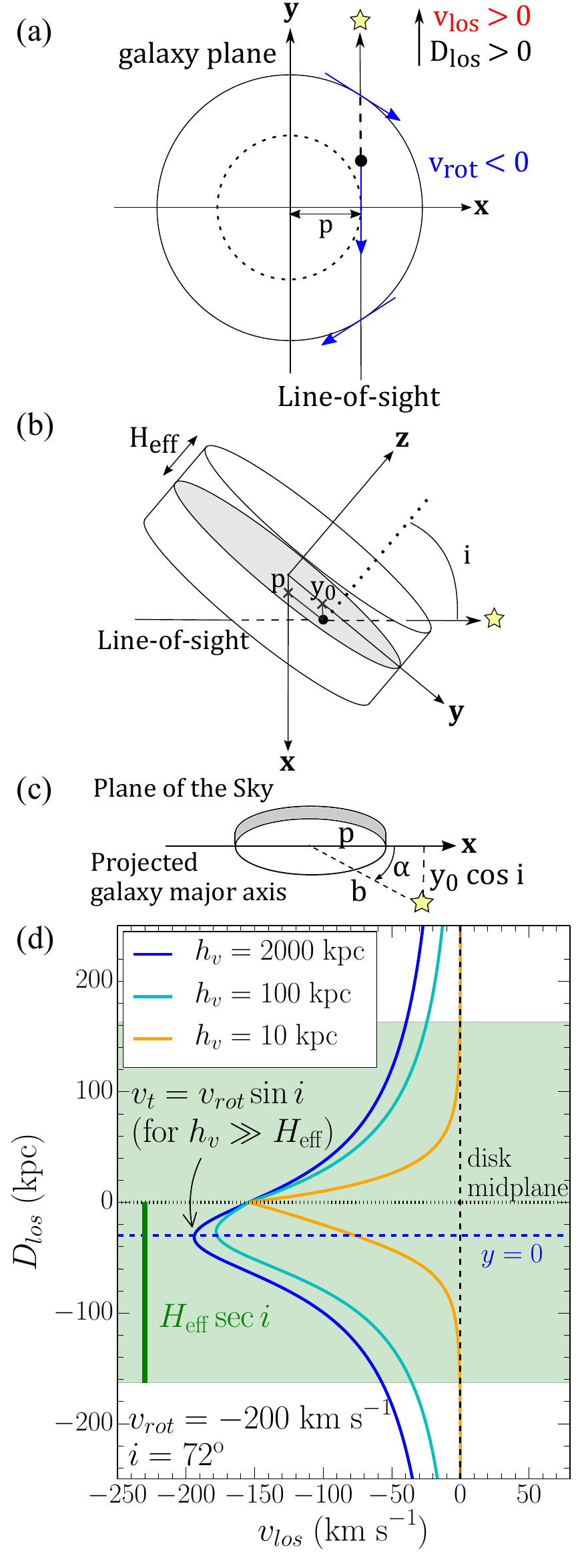}
	\caption{
			Rotating disk model.
			The four diagrams show the setup of the coordinate system and the graphical representation
			of each variable, similar to those in \citet{Steidel2002} except for 
			the direction of the sightline.  
			These panels also show the sign conventions of line-of-sight velocity $v_{los}$ and  
			rotation velocity $v_{rot}$, with red and blue arrows indicate 
			whether redshifted or blueshifted velocities are detected.  
			(a) The plane of the galactic disk.  
			(b) Side view of the disk.  
			(c) The galaxy projected onto the sky.  
			(d) The position--velocity diagram.  
			The line-of-sight velocity component is plotted as a function of distance along the 
			quasar sightline.  The green shaded region indicates the portion of the sightline 
			intersecting the disk.  
			Decreasing the velocity scale height $h_v$ from 2000 to 10 kpc moves the 
			maximum Doppler shift along $D_{los}$ from $y=0$ toward the disk midplane.
			}
	\label{fig:rot_vert}
\end{figure}

To interpret the data we have presented in this paper, we describe a simple model 
introduced by \citeauthor{Steidel2002}~(\citeyear{Steidel2002}, hereafter S02).\footnote{
       We change the sign convention. 
	   We adopt a sign convention for which gas moving toward (away from) the observer
	   will give a negative (positive) line-of-sight velocity. 
	   See Equations~(\ref{eq:sk-rotation})--(\ref{eqn:v_los_sk}) and Figure~{\ref{fig:rot_vert}}.	   
       }
The simple model extends the galactic disk far
enough in radius that the quasar sightline intersects it.
The inclination of the disk and its rotation speed, which we assume
are independent of radius, are well constrained by observations of 
the galaxy.
To address the failures of this description,
we introduce a radial infall model
motivated by simulations that show recently accreted gas 
near the plane of the galactic disk \citep{Stewart2011b,Stewart2013,Danovich2015}.
The full model combines these components
and is described by four parameters that can be varied in
order to change the velocity range of the absorption: 
the disk thickness $H_\mathrm{eff}$, the velocity scale height $h_v$,
the radial velocity $v_R$, 
and the sign of the inclination, which describes the orientation of the disk flip.

\subsection{Kinematic Signatures of Circular Orbits in Disks}
\label{sec:model}

We adopt a right-handed coordinate system 
with the origin at the center of the galaxy.  
The $z$-axis is perpendicular to the disk plane, i.e.,  
the $xy$-plane as illustrated in Figure~\ref{fig:rot_vert}.  
Without the loss of generality, we place the sightline in the $x = p$ plane. 
When projected onto the sky, the $x$-axis is aligned with 
the major axis of the galactic disk, and the length of the impact parameter 
projected onto the galaxy major axis is $p$. We 
define $y_0$ as the $y$-coordinate of the sightline where it intersects 
the midplane of the disk.  The geometry in Figure~\ref{fig:rot_vert}
shows that a sightline at azimuthal angle $\alpha$ and impact parameter $b$
is described by
\begin{equation}
	p = b \cos \alpha \quad \mathrm {and} \quad y_0 \cos i = b \sin \alpha
	~.
	\label{eq:balpha-link}
\end{equation}

This geometry allows two orientations of the disk, which are indistinguishable
from galaxy morphology alone. We select the positive direction of the $z$-axis to 
always lie on the quasar's side of the disk. When the $y$-coordinate of the quasar is 
positive, as shown in Figure~\ref{fig:rot_vert}, we give the inclination of the disk
a positive value. To describe the situation where the disk is rotated 45\deg\ about
the $x$-axis, and the $y$-coordinate of the quasar becomes negative, we introduce the
concept of a negative inclination. From the perspective of an observer, this
{\it flip} about the major axis corresponds to interchanging the near and far sides
of the disk. For circular orbits in the disk plane, the primary  observable
signature of this geometrical {\it flip} is the wrapping direction of trailing
spiral arms relative to the direction of rotation.  

The thickness of the disk, $H_\mathrm{eff}$ in Figure~\ref{fig:rot_vert}, and the 
vertical gradient in rotation speed can be varied to model the data. We describe 
the $|z|$-dependence of the rotation speed with the exponential velocity scale 
height, $h_v$, adopted by S02. Projection of this thick rotating disk,
\begin{equation}
	\boldsymbol{v}_\phi(z) = v_{rot}  \exp(-|z|/h_v) \boldsymbol{\hat{\phi}}
        ~,
	\label{eq:sk-rotation}
\end{equation}
onto the vector describing the quasar sightline,
\begin{equation}
	\boldsymbol{L} = (\sin i) \boldsymbol{\hat{y}} + (\cos i) \boldsymbol{\hat{z}}
	~,
	\label{eq:sightline}
\end{equation}
yields line-of-sight velocities
\begin{equation}
	v_{los} = \boldsymbol{v}_\phi \cdot \boldsymbol{L} = 
		\frac{v_{rot} \sin i}{\sqrt{1 + (y/p)^2}} \exp \left(-\frac{|y-y_0|}{h_v \tan i}\right)
	~.
	\label{eqn:v_los_sk}
\end{equation}
With the right-handed coordinate system, 
the rotation velocity $v_{rot}$ is positive (negative) when
the angular momentum vector aligns with the positive (negative) 
$z$-axis of the disk.   
Panel (d) of Figure~\ref{fig:rot_vert} displays $v_{los}$ along the sightline
defined by
\begin{equation}
	D_{los} = (y - y_0) / \sin i
	~,
	\label{eq:d-los}
\end{equation}
such that $D_{los} = 0$ represents the midplane of the disk.

The disk thickness $H_\mathrm{eff}$ sets boundaries on the range of $y$-values, 
\begin{equation}
	y_0 - H_\mathrm{eff}\tan i < y < y_0 + H_\mathrm{eff}\tan i
	~,
	\label{eq:y-bound}
\end{equation}
within the disk.  
The green shaded region in Figure~\ref{fig:rot_vert}(d) indicates 
the portion of the quasar sightline within the disk.  
Increasing the $H_\mathrm{eff}$ 
broadens the range of $v_{los}$ from the projected
rotation speed at the point where the sightline is tangent to the disk,
$v_t$, toward an asymptotic value $v_{los} \rightarrow\ 0$.

As suggested by S02,
we introduce a  velocity scale height $h_v$ to govern how quickly the rotation 
velocity drops from $v_t$ toward zero with distance along the sightline.  
Choosing $h_v >> H_\mathrm{eff}$
produces a linear decline in rotation speed with height above the disk
\begin{equation}
	\boldsymbol{v}_\phi(z) \approx v_{rot} (1 -|z|/h_v) \boldsymbol{\hat{\phi}}
        ~.
\end{equation}
As a fiducial reference point, we adopt $h_v = 10$~kpc, which produces
a vertical rotation  gradient  consistent with measurements for extraplanar
gas above nearby galaxies \citep{Benjamin2002,Benjamin2012}.

The position--velocity  ($D_{los}$ vs.\ $v_{los}$)  diagrams illustrate 
the variation in line-of-sight velocity along a particular sightline.  
For example, consider a very large velocity scale height, 
which effectively describes cylindrical rotation. In the limit
$\lim h_v \rightarrow \infty$, inspection of Equation~(\ref{eqn:v_los_sk}) shows
that the maximum Doppler shift is contributed by gas at $y = 0$. 
The sightline is tangent to the disk at the point $y = 0$, and the maximum 
line-of-sight velocity becomes
\begin{equation}
  v_t \equiv v_{los}(y=0) = v_{rot} \sin i
  ~.
\end{equation}
The vertical velocity gradient steepens with decreasing $h_v$, 
which moves the maximum Doppler shift along $D_{los}$ from $y = 0$ toward the disk midplane.  
The three curves in Figure~\ref{fig:rot_vert}(d) demonstrates 
this shift when $h_v$ decreases from 2000 kpc to 10 kpc.  
The exponential term in Equation~(\ref{eqn:v_los_sk}) quantitatively describes the shift.

\subsection{Kinematics Signatures of Radial Infall in the Disk Plane}
\label{sec:model-vradial}

\begin{figure*}[bth]
	\centering
	\includegraphics[width=0.9\linewidth, trim = 0 -40 0 0]{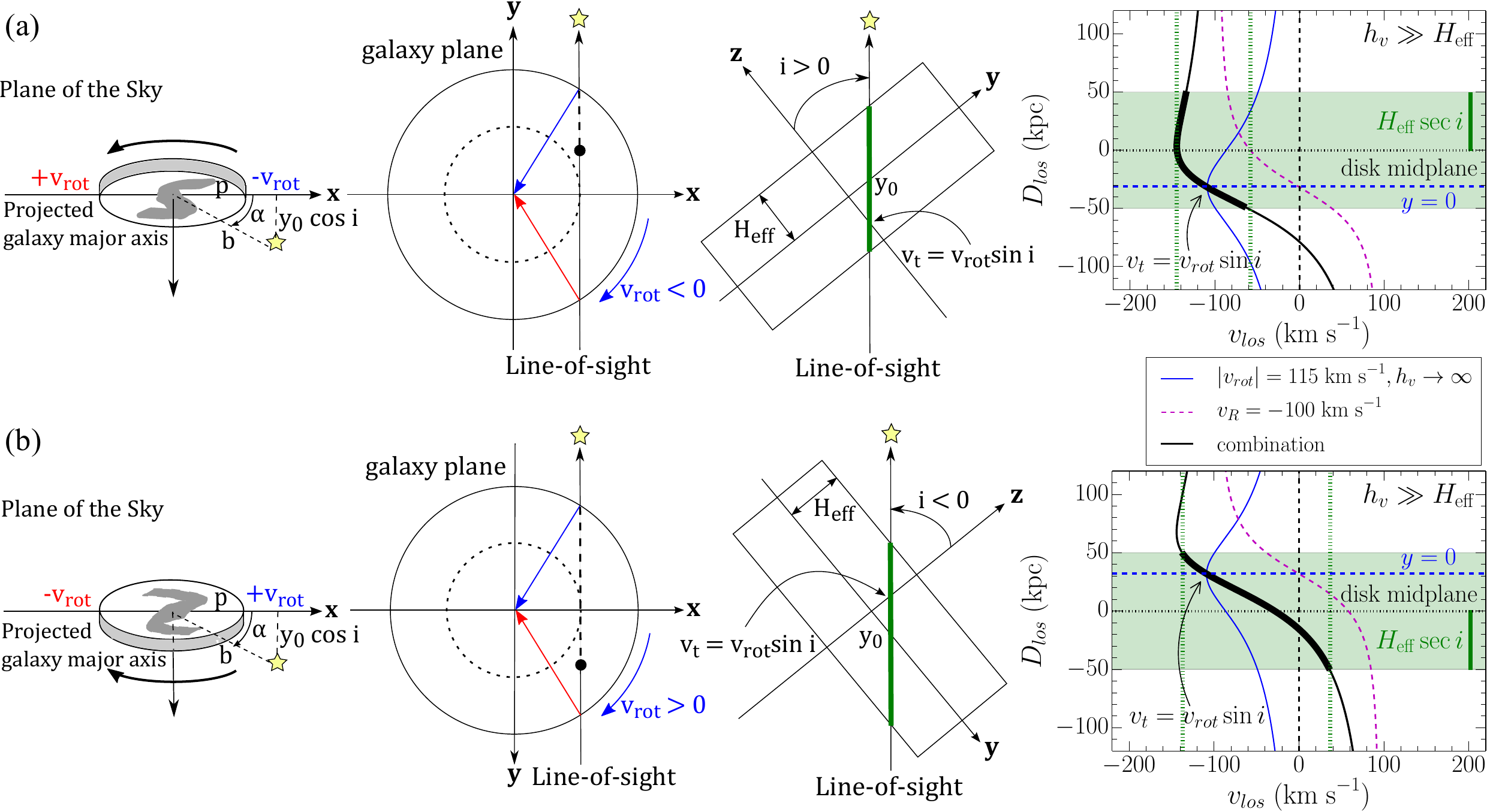}
	\caption{Radial infall in the disk plane plus rotation. The radial infall
          contributes redshifted (blueshifted) 
          absorption on the near (far) side of the sightline's passage through the disk.
          For a disk with a measured $b/a$-ratio, the two possible tilts of the disk
          with respect to the plane of the sky are shown; we label these as positive 
          and negative inclinations for clarity.  
		  In the first two columns, $v_{rot}$ changes sign between the two tilts
		  even a blueshifted rotation component is detected along each quasar sightline; 
		  this is because the coordinate system is right-handed.
          The third column shows that the disk tangent point falls on opposite sides
          of the disk midplane for the two tilts. The $D_{los} - v_{los}$ curves
          shift in the last column. Since the height of the disk ($H_{\mathrm{eff}}$)
          is symmetric with respect to the midplane, the portion of the sightline
          with gas shifts with the disk tilt.  The two tilts produce absorption
          over a different range of velocity (enclosed by vertical green dashed lines) 
		  when inflow is included.  
		  }
	\label{fig:disk_ori}
\end{figure*}

We add a velocity component so that the gas motion in the disk plane becomes
\begin{equation}
	\boldsymbol{v}(R,z) = 
		\boldsymbol{v}_R(R,z) \boldsymbol{\hat{R}} + \boldsymbol{v}_\phi(R,z) \boldsymbol{\hat{\phi}}
	~.
	\label{eq:add-vr}
\end{equation}
For purposes of illustration, we consider only constant infall speeds
such that
\begin{equation}
	\boldsymbol{v}_R(R,z) = v_R \boldsymbol{\hat{R}}
	~,
	\label{eq:vr}
\end{equation}
where $v_R < 0$ for infalling gas.  

Similar to Equation~(\ref{eqn:v_los_sk}), the line-of-sight velocity due to 
this radial velocity component {\it alone} is 
\begin{equation}
	v_{los,R} = \boldsymbol{v}_R \cdot \boldsymbol{L} = (v_R \sin i) \frac{y}{\sqrt{y^2 + p^2}}
	~,
	\label{eq:v-los-vr}
\end{equation}
which gives a total line-of-sight velocity of
\begin{equation}
	v_{los} = v_{los,\phi} + v_{los,R}
	~.
	\label{eq:v-los-sk-r}
\end{equation}

The line-of-sight velocity of the radial infall changes sign along the sightline 
as illustrated by Figure~\ref{fig:disk_ori}.
In the last column, the line-of-sight velocity from the rotation component
lies entirely to one side of the systemic velocity while the infall component
changes sign. The inflection point for the infall velocity corresponds to the
point where the sightline is tangent to the disk, and the projected rotation component
reaches a maximum.

Flipping the disk changes how the radial and azimuthal vector velocities add together.
Only in the limit of large $H_\mathrm{eff}$ will the velocity ranges of the two
orientations be identical. To our knowledge, this asymmetry has not been previously
described.  
In the last column of Figure~\ref{fig:disk_ori}, 
we illustrate how the boundaries imposed by the disk thickness
can lead to only one orientation producing absorption that spans the systemic
velocity. 
We indicate the velocity range of absorption using a pair of green dashed lines.  
In our example, only the disk tilt with negative inclination 
produces absorption across the systemic velocity.

Another interesting situation is that the orientation of the flip moves the location 
of the maximum Doppler shift along the sightline.  
In our example in Figure~\ref{fig:disk_ori}, when the disk inclination is positive, 
the maximum Doppler shift is contributed by gas near the disk midplane.  
Flipping the disk, however, moves the maximum Doppler shift away from the midplane.
When the model includes radial inflow, the sign of the disk inclination therefore changes 
the kinematics of an absorption-line system.

\end{document}